\documentclass[a4paper,fleqn]{cas-sc}

\usepackage[authoryear,longnamesfirst]{natbib}

\def\tsc#1{\csdef{#1}{\textsc{\lowercase{#1}}\xspace}}
\tsc{WGM}
\tsc{QE}
\tsc{EP}
\tsc{PMS}
\tsc{BEC}
\tsc{DE}

\begin{document}
\let\WriteBookmarks\relax
\def\floatpagepagefraction{1}
\def\textpagefraction{.001}

\shorttitle{Functional neural network control chart}

\shortauthors{Kulahci et~al.}

\title [mode = title]{Functional neural network control chart}                      



%
\author[1,2]{Murat Kulahci}[orcid=0000-0003-4222-9631]



\ead{muku@dtu.dk}



\affiliation[1]{organization={Technical University of Denmark, Department of Applied Mathematics and Computer Science},
    city={Kongens Lyngby},
    country={Denmark}}

\affiliation[2]{organization={Luleå University of Technology, Department of Business Administration, Technology and Social Sciences},
    city={Luleå},
    country={Sweden}}

\author[3]{Antonio Lepore}[orcid=0000-0002-8739-5310]

\cormark[1]


\ead{antonio.lepore@unina.it}


\author[3]{Biagio Palumbo}[orcid=0000-0003-1036-8127]
\ead{biagio.palumbo@unina.it}

\author[3]{Gianluca Sposito}[orcid=0000-0002-4885-1635]
\ead{gianluca.sposito@unina.it}


\affiliation[3]{organization={University of Naples Federico II, Department of Industrial Engineering},
    city={Naples},
    country={Italy}}



\cortext[cor1]{Corresponding author}



\begin{abstract}
In many Industry 4.0 data analytics applications, quality characteristic data acquired from manufacturing processes are better modeled as functions, often referred to as profiles. 
In practice, there are situations where a scalar quality characteristic, referred to also as the \textit{response}, is influenced by one or more variables in the form of functional data, referred to as \textit{functional covariates}.
To adjust the monitoring of the scalar response by the effect of this additional information, a new profile monitoring strategy is proposed on the residuals obtained from the functional neural network, which is able to learn a possibly nonlinear relationship between the scalar response and the functional covariates. 
An extensive Monte Carlo simulation study is performed to assess the performance of the proposed method with respect to other control charts that appeared in the literature before.  
Finally, a case study in the railway industry is presented with the aim of monitoring the heating, ventilation and air conditioning systems installed onboard passenger trains.
\end{abstract}




\maketitle

\section{Introduction}

In many modern production processes, advanced data acquisition systems generate massive amounts of data in the form of curves or surfaces varying over a continuum, such as time or space. Such data can be best modeled as functions that are more generally defined multidimensionally, and are usually referred to as \textit{profiles} or \textit{functional data} \citep{ramsaybook, menafoglio2017statistical, hsing2015theoretical, ferraty2006nonparametric}. 
The statistical monitoring of a process that is best characterized by functional data is known as profile monitoring.
Some examples of profile monitoring applications can be found in the works of \cite{jin1999feature, woodall2004using, zou2007monitoring, williams2007statistical, colosimo2010comparison, saghaei2013statistical}. 
As an application area within statistical process monitoring (SPM) \citep{stuart1996statistical}, profile monitoring aims at the detection of special causes of variation acting on the process, which, in such cases, is said to be out of control (OC). Otherwise, the process is said to be affected by only random causes of variation and hence in control (IC). 
More specifically, profile monitoring focuses on testing the stability of the functional relationship between the quality characteristic of interest, also referred to as the response, and one or more exploratory variables, referred to as covariates.
That is, the quality characteristic of interest is monitored conditionally on the covariate levels.
The regression control chart \citep{mandel1969regression} is the first scheme proposed in statistical literature to address this issue by monitoring the residuals of the regression of the quality characteristic on the covariates.
In this way,  the monitoring strategy is able to consider the variance explained by the covariates and to leverage their additional information to improve the power of the monitoring scheme expressed as the probability of correctly detecting a shift in the process being monitored \citep{montgomery2020introduction}.
Recent and more advanced extensions of this idea allow the covariates or the response to be described by functional data themselves, such as the functional regression control chart (FRCC) framework proposed by \cite{centofanti2021functional} where functional models \citep{morris2015functional} can be used to map the influence of one or more functional covariates on a scalar or functional response. 
However, all the instances that can be reconducted to the FRCC framework, such as \cite{capezza2020control, capezza2022functional,capezza2022robust, centofanti2022real}, are implemented by means of functional \textit{linear} models, only. 

In the meantime, neural networks (NNs) and deep learning (DL) techniques have been receiving increasing attention in time series, computer vision, speech recognition, and genetics. 
However, despite their success, their use for functional data is not completely investigated.
\cite{rossi2005representation} discussed how to incorporate functional pre-processing, e.g., functional principal component analysis, into a multilayer perceptron (MLP), which is a particular type of NN.
\cite{rossi2002functional,conan2002multi} introduced the functional multilayer perceptron (FMLP), extending the MLP to functional data through a proper reparameterization of the weight matrices. 
In particular, the former proposes an FMLP that simply requires a discretization of the functional covariates and, thus it cannot directly handle profiles with a different number of observations or measured at different time points. 
To overcome this issue, the latter uses the basis expansions \citep{wang2016functional} of the functional covariates.
The theoretical properties of the FMLP are firstly studied by \cite{rossi2006theoretical}, where the authors revealed that an FMLP is a universal approximator following the definition of  \cite{hornik1989multilayer}.
FMLP has also been further investigated by \cite{wang2019remaining, wang2019multilayer}.

Another common approach to modeling functional data is through convolutional (CNN) \citep{lecun1998gradient} and recurrent NN (RNN) \citep{rumelhart1986learning}. In particular, long short-term memory variant \citep{hochreiter1997long} is the most popular, due to the ability to recognize patterns for a long duration of time. 
However, these may fail to learn the underlying smoothness of the functional data from raw noisy measurements, which are very common in real-data applications.
\cite{yao2021deep} proposed an adaptive functional NN (AdaFNN) that is a NN specifically designed for functional data through the definition of a new \textit{basis} layer which implements a micro NN \citep{lin2013network} to directly learn the most relevant base functions to represent the response value, thus avoiding a pre-specified choice.
The authors demonstrate the superior performance of their proposal over the MLPs through an extensive simulation study, but they do not compare their proposal with any of the functional regression models already present in the functional data (FDA) literature. 
Additionally, even though the authors claimed that the functional coefficients learned by AdaFNN are interpretable, it is not true in general as each micro NN learns a different functional coefficient and it is impossible to priorly know how to combine them and resemble the true functional coefficients.
Recently, \cite{thind2023deep} introduced a new DL architecture, known as functional neural network (FNN), which allows for deep architectures for scalar responses with multiple functional and scalar covariates through the definition of smooth weight functions.
The \textit{functional} weight can be visualized during the training, increasing the interpretability of the FNN while maintaining the nonlinear predictive power of traditional NNs.
Furthermore, the number of parameters of the FNN can be lower than the amount needed in traditional MLPs, CNNs, and RNNs.(see Section \ref{sec:Functional Neural Network} for more details). 
Through simulated and benchmark data sets, \cite{thind2023deep} demonstrated that FNN outperforms state-of-the-art DL architectures, functional linear models, and a number of other multivariate methods, e.g., least square regression and random forest, in terms of prediction accuracy \citep{hastie2009elements}.
FNN is also proven to be a universal approximator \citep{cybenko1989approximation}, that is it can be used to learn any continuous function to a desired degree of accuracy. 
In recent years, different DL-based have been proposed for the SPM of multivariate processes and functional data, and some relevant examples can be found in \cite{stuart1996statistical, sergin2021toward, pacella2011monitoring, chen2020monitoring, howard2018identifying, YEGANEH2022117572, YEGANEH2021114237}.

In this paper, we propose a novel control chart, named FNN control chart (FNNCC), that exploits the FNN architecture advantages for the monitoring of a scalar quality characteristic when functional and scalar covariates are available.
The proposed FNNCC can be regarded as an implementation of the FRCC framework where the influence of functional or scalar covariates on a scalar response does not need to be necessarily linear as in \cite{capezza2020control, capezza2023functional}, which is based on a linear scalar-on-function (SOF) regression model \citep{reiss2017methods}.

The article is structured as follows. 
Section \ref{sec: Methodology} introduces some background on the SOF regression model and FNNs. Then, we describe the proposed FNNCC in detail. 
An extensive Monte Carlo simulation study is performed in Section \ref{sec: Simulation study} to quantify the FNNCC OC average run length ($ARL_1$) at a given IC average run length ($ARL_0$) \citep{qiubook} in identifying a mean shift in the scalar response in the presence or absence of drifts in the covariate mean and to compare it with other competing control charting schemes that have already appeared in the literature before. 
The practical applicability of the proposed method is illustrated in Section \ref{sec: Real-Case Study} by means of a case study in the monitoring of heating, ventilation and air conditioning (HVAC) systems installed on passenger trains \citep{lepore2022neural}.
The \texttt{HVAC} data set, courtesy of the rail transportation company \textit{Hitachi Rail Italy}, and the analysis code are available online at \url{https://github.com/unina-sfere/FNNCC}. 
The final section contains concluding remarks and outlines new directions for future research.
Additionally, in Appendix A, we implement two other NN-based monitoring strategies and compare them to the proposed control chart.
All computations and plots have been obtained using the programming language R \citep{R}.



\section{Methodology}\label{sec: Methodology}

In this section, we briefly review the SOF linear regression model in Section \ref{sec:FDA}. Then, we introduce the FNN in Section \ref{sec:Functional Neural Network} and provide details of the monitoring strategy based on the FNN in Section \ref{sec:The monitoring strategy}.

\subsection{Functional linear regression}\label{sec:FDA}
We begin by providing the necessary notation used in the article, followed by a summary of the linear model for SOF regression before describing the proposed methodology.

\noindent Let $y_i$ and $\boldsymbol{X}_i=\left\{\left(X_{i1}, \ldots, X_{iP}\right)^{\top}\right\}$, $i=1, \ldots, n$, denote $n$ observations of a scalar response covariates and a vector of $P$ functional variables, respectively.
$\boldsymbol{X}_i$ is a random element that takes values in the Hilbert space $L^2(\mathcal{T})^P$, i.e., $X_{i,1}, \ldots X_{i,P}$ belong to $L^2(\mathcal{T})$, the space of square-integrable functions defined on the compact interval $\mathcal{T}$.  
We also assume that $\mathbf{X}_i$ is fully observed \citep{kokoszka2017introduction} i.e., is densely observed on a set of discrete grid points. 
The aim is to learn the mapping $G: \mathcal{L}^2(\mathcal{T}) \times$ $\ldots \times \mathcal{L}^2(\mathcal{T}) \rightarrow \mathcal{R}$ from the functional covariates $\mathbf{X}_i$ to the scalar response $y_i$
\begin{equation} 
\label{regression model}
y_i =G\left(\mathbf{X}_i\right) + \epsilon_i \quad i=1, \ldots, n, 
\end{equation}
where $\epsilon_i$ is a scalar error term.
The SOF regression has been extensively studied in the FDA literature and, in this section, we give a succinct description of the SOF linear regression model used by \cite{capezza2020control}, defined as 
\begin{equation}
\label{eqn:sof linear model}
y_i =\alpha+\sum_{p=1}^P \int_{\mathcal{T}} \beta_p(t) X_{ip}(t) d t + \epsilon_i \quad i=1, \ldots, n,
\end{equation}
where $\alpha \in \mathbb{R}$ is the scalar intercept, $\boldsymbol{\beta}=\left(\beta_1, \ldots, \beta_p, \ldots, \beta_P\right)^{\top} \in L^2(\mathcal{T})^P$ are the functional coefficients to be estimated and $\epsilon_i$ are the error terms, which are assumed to be independent and identically distributed normal random variables with mean zero and variance $\sigma^2$.
Comparing Equation \eqref{regression model} and Equation \eqref{eqn:sof linear model}, it is trivial to note that in the linear SOF regression model $G\left(\mathbf{X}_i\right) = \alpha+\sum_{p=1}^P \int_{\mathcal{T}} \beta_p(t) X_{ip}(t) d t$.
Without loss of generality, the functional covariates are assumed to be empirically standardized \citep{chou2014monitoring}, that is each covariate observation is standardized by subtracting pointwise the corresponding sample mean and dividing the result by the relative standard deviation function. 
The coefficients $\alpha$ and $\boldsymbol{\beta}$ in Equation \eqref{eqn:sof linear model} can be estimated by solving the following minimization problem
\begin{equation}
\label{eqn: least-squares problem}    
\min _{\alpha \in \mathbb{R}, \boldsymbol{\beta} \in L^2(\mathcal{T})^P} \sum_{i=1}^n\left(y_i - \alpha - \sum_{r=1}^P \int_{\mathcal{T}} \beta_p(t) X_{ip}(t) d t\right)^2.
\end{equation}
Because of the infinite dimensionality of the functional data, the above minimization problem is not well-posed and the model cannot be estimated using the least squares approach \cite{james2013introduction} directly. 
However, being square integrable, the functional covariates can be represented through the \textit{Karhunen-Loéve expansion} as follows 
\begin{equation}
\label{eqn: Karhunen-Loéve expansion}
\boldsymbol{X}_{i}(t)=\sum_{m=1}^{\infty} \xi_{im} \boldsymbol{\psi}_{m}(t), \quad t \in \mathcal{T}, \quad i=1 \ldots, n,
\end{equation}
where the $\boldsymbol{\psi}_m=\left(\psi_{m1}, \ldots, \psi_{mP}\right)_{m \in \mathbb{N}} \in L^2(\mathcal{T})^P$ are the multivariate functional principal components (MFPCs) defined as the eigenfunctions of the covariance function $\mathbf{c}(s, t)$ of the multivariate functional data i.e. they are the solutions to the equation
\begin{equation}
\label{eqn: eigen problem}
\int_{\mathcal{T}} \mathbf{c}(s, t) \boldsymbol{\psi}_m(s) d s=\lambda_m \boldsymbol{\psi}_m(t), \quad \forall t \in \mathcal{T},
\end{equation}
where $\lambda_m$ are called the eigenvalues of $\mathbf{c}(s, t)$. 
The eigenvalues $\lambda_m$, and so the corresponding eigenfunctions $\boldsymbol{\psi}_m(t)$, are arranged in non-increasing order $\lambda_1 \geq \lambda_2 \geq \cdots \geq 0$. The eigenfunctions $\boldsymbol{\psi}_m(t)$ are by construction such that 
\begin{equation}
\label{orthonormal basis}
\sum_{p=1}^{P} \int_{\mathcal{T}} \psi_{m_1,p}(t) \psi_{m_2,p}(t) d t = 
\begin{cases} 1, & \mbox{if } m_1 = m_2 \\ 0, & \mbox{if }  m_1 \neq m_2,
\end{cases}
\end{equation}
that is, they form an orthonormal basis of $L^2(\mathcal{T})^P$. The terms $\xi_{im}$ in Equation \eqref{eqn: Karhunen-Loéve expansion} are called the MFPC scores, or simply scores, and defined as 
\begin{equation}
\label{eqn: functional score}
\xi_{im} = \sum_{p=1}^{P} \int_{\mathcal{T}} X_{ip}(t) \psi_{mp}(t) d t \quad i=1 \ldots, n.
\end{equation}
It can be shown that $\text{E}(\xi_{m}) = 0$, $\text{E}(\xi_{m}^2) = \lambda_m$ and $\text{E}(\xi_{m_1},\xi_{m_2}) = 0$ when $m_1 \neq m_2$.
The decomposition in Equation \eqref{eqn: Karhunen-Loéve expansion} is optimal in the sense that, for each finite $M \in N$, $\boldsymbol{X}_{i}(t)$ is best approximated by $\mathbf{\hat{X}}_i = (\hat{X}_{i1}, \ldots, \hat{X}_{ip},\ldots ,\hat{X}_{iP})^T$, obtained by truncating the Karhunen-Loéve expansion, as follows  
\begin{equation}
\label{eqn: expansion approximation}
\hat{X}_{ip}(t)=\sum_{m=1}^{M} \xi_{im} {\psi}_{mp}(t), \quad t \in \mathcal{T}, \quad p=1,\ldots,P, \quad i=1 \ldots, n.
\end{equation}
It can be proved that, upon using $\mathbf{\hat{X}}_i$ in place of $\mathbf{X}_i$, the functional coefficient $\boldsymbol{\beta}$ in Equation \eqref{eqn:sof linear model} can be replaced with $\hat{\boldsymbol{\beta}}=\left(\hat{\beta}_{1}, \ldots, \hat{\beta}_{p}, \ldots, \hat{\beta}_{P}\right)^{\top} \in L^{2}(\mathcal{T})^{P}$ by using the same truncated basis expansion, that is
\begin{equation}
\label{eqn: beta regression}    
\hat{\beta}_{p}(t)=\sum_{m=1}^{M} b_{m} \psi_{mp}(t), \quad t \in \mathcal{T}, \quad p=1, \ldots, P,
\end{equation}
where $b_{1}, \ldots, b_{M}$ are the basis coefficients. 
Upon using the approximation in Equations \eqref{eqn: expansion approximation} and Equation \eqref{eqn: beta regression}, Equation \eqref{eqn:sof linear model} can be rewritten as 
\begin{equation}
\label{eqn: sof model least square}
y_{i}=\alpha +\sum_{m=1}^{M} \xi_{im} b_{m}+\varepsilon_{i}^{*}, \quad i=1, \ldots, n.
\end{equation}
Hence, in this form, the model parameters, namely the intercept $\alpha$ and the coefficient $b_m$, $m = 1, \ldots, M$, can be estimated by using the least square approach, as $\hat{\alpha}=\frac{1}{n} \sum_{i=1}^{n} y_{i}$ and $\hat{b}_{m}=\sum_{i=1}^{n} y_{i} \xi_{i m} / \sum_{i=1}^{n} \xi_{i m}^{2}$, respectively.
Accordingly, the least square prediction of the scalar response $y_i$ can be obtained as
\begin{equation}
\label{eqn: sof prediction}
\hat{y}_{i}=\hat{\beta}_{0}+\sum_{p=1}^{P} \int_{\mathcal{T}} X_{ip}(t) \hat{\beta}_{p}(t) d t, \quad i=1, \ldots, n
\end{equation}

As in the multivariate setting, the number $M$ can be chosen such that the retained MFPCs $\boldsymbol{\psi}_m=\left(\psi_{m1}, \ldots, \psi_{mP}\right)_{m = 1, \ldots, M}$ explain at least a given percentage, say 80\%, of the total variability. 
In this paper, we use a different strategy based on the final model's predictive ability.
That is, the $M$ MFPCs to be retained are the first $M$ that achieve a given reduction in the prediction sum of squares statistic, defined as $\sum_{i=1}^{n}\left(y_{i}-\hat{y}_{[i]}\right)^{2}$, where $\hat{y}_{[i]}$ is the $i$-th fitted value of the scalar response based on the SOF regression model with the $i$-th observation removed from the data set used to fit the linear model. In this way, this strategy ensures that the MFPCs with a small predictive ability are not retained in the SOF model.
More details on this problem can be found in \cite[p.~173-177]{jolliffe2016principal}.

\subsection{Functional neural network}\label{sec:Functional Neural Network}

NNs are computational models inspired by the structure and functioning of the human brain and consisting of interconnected artificial neurons, also known as nodes or units. The latter are organized into layers, typically an input layer, one or more hidden layers, and an output layer. 
The input layer receives the initial input data, and the output layer produces the final output or prediction. The hidden layers are intermediate layers between the input and output layers and play a crucial role in learning complex representations of the data.
Each neuron takes multiple inputs from the previous layer, performs a weighted sum, and then applies a nonlinear function to produce an output. 
Let $n^{(j)}$ and $\mathbf{h}^{(j)}$ be the number of neurons in the $j$th hidden layer and the output of the $j$th hidden layer, respectively.
More formally, $\mathbf{h}^{(j)}$ is defined as $\mathbf{h}^{(j)}=g\left(\mathbf{W}^{(j)} \boldsymbol{h}^{(j-1)}+\boldsymbol{b}^{(j)}\right)$, where 
$\mathbf{h}^{(j-1)}\in R^{n^{(j-1)}}$ represents the output of the previous $(j-1)$th layer, $\mathbf{W}^{(j)}$ is a $n^{(j)} \times n^{(j-1)}$ weight matrix and $\boldsymbol{b}^{(j)} \in R^{n^{(j)}}$ is the intercept, often referred to as the bias in the machine learning field.
The function $g: \mathbb{R}^{n^{(j)}} \rightarrow \mathbb{R}^{n^{(j)}}$ is called the activation function and introduces nonlinearity into the output of the neuron \citep{hastie2009elements}, which may enable more complex pattern recognition and more accurate predictions. 
If $g$ is the identity function, any NN is proved to specialize into a linear regression model. 
The choice of the activation function depends on the type of problem to be solved. 
Some common ones are (a) the sigmoid function (or logistic function) \citep{han1995influence} that maps the input to a range between 0 and 1, making it useful for binary classification problems; (b) the rectified linear unit (ReLU) \citep{hahnloser2000digital} that returns the input if it is positive, and zero otherwise and is widely used due to its simplicity and computational efficiency; (c) 
the hyperbolic tangent \citep{rumelhart1985learning} that squashes the input values to the range between -1 and 1, making it useful for classification tasks; and (d) the softmax function \citep{ackley1985learning} that is commonly used in the output layer of a NN for multi-class classification problems, as it converts the outputs into a probability distribution.

Traditional NNs accept only finite-dimensional vectors as input and thus, they cannot easily handle profiles, whereas, the FNN introduced by \cite{thind2023deep} is instead purposely designed for it.

\noindent Given $X_1(t), \ldots, X_P(t)$ and $z_1, \ldots, z_J$ functional and scalar covariates, respectively, \cite{thind2023deep} introduce the functional weights $\boldsymbol{\gamma}(t)=\left(\gamma_{1}, \ldots, \gamma_{p}, \ldots, \gamma_{P}\right)^{\top}$ to effectively weigh the functional covariates at every point along their domain $\mathcal{T}$, and define the output $h_k^{(1)}$ of the $k$th neuron in the first hidden layer corresponding to the $i$th observation as follows
\begin{equation}
\label{eqn: FNN first layer}
h_{ik}^{(1)}=g\left(\sum_{p=1}^P \int_{\mathcal{T}} \gamma_{kp}(t) X_{ip}(t) d t+\sum_{j=1}^J w_{kj}^{(1)} z_{ij}+b_k^{(1)}\right) \quad i=1,\ldots,n \quad k = 1,\ldots, n^{(1)}, 
\end{equation}
where $n^{(1)}$ is the number of neurons in the first hidden layer and $g(\cdot)$ is the activation function.
This layer is referred to as a \textit{functional} hidden layer as it consists of neurons capable of handling infinite dimensional \textit{functional} weights $\boldsymbol{\gamma}(t)$.
Each neuron in the first layer produces a scalar value that is then fed into a regular NN. 
It is worth noting that only the output of the first hidden layer has this \textit{functional} structure and this is the reason why $\gamma_{kp}(t)$ in Equation \eqref{eqn: FNN first layer} does not need for any superscript.
This implies that the rest of the hidden layers of the FNN can be of any of the usual forms (e.g., feedforward, recurrent, residual) \citep{lecun2015deep}.
Similarly to the results in Equation \eqref{eqn: beta regression}, the functional weights ${\gamma_{kp}}(t)$ in this first hidden layer can be approximated through a linear combination of basis functions
\begin{equation}
\label{eqn: beta expansion}
\hat{\gamma}_{kp}(t)=\sum_{m=1}^{M_p} c_{kpm} \zeta_{kpm}(t)=\boldsymbol{c}_{kp}^T \boldsymbol{\zeta}_{kp}(t) \quad p=1,2, \ldots, P \quad k = 1,2, \ldots, n^{(1)}, 
\end{equation}
where $\boldsymbol{\zeta}_{kp}(t)=\left(\zeta_{kp1}(t), \ldots, \zeta_{kpM_p}(t)\right)^T$ is a vector of basis functions,  $\boldsymbol{c}_{kp}=\left(c_{kp1}, \ldots, c_{kpM_p}\right)^T$ is the corresponding vector of basis coefficients to be estimated by the NN, and 
$M_p$ denotes the number of basis functions for each of the $P$ functional covariates.
Using the basis approximation in Equation \eqref{eqn: beta expansion}, the general form of the $k$th neuron in the first hidden layer in Equation \eqref{eqn: FNN first layer} can be rewritten as
\begin{equation}
\label{eqn: FNN general first layer}
\begin{aligned}
h_k^{(1)} & =g\left(\sum_{p=1}^P \int_{\mathcal{T}} \sum_{m=1}^{M_P} c_{kpm} \zeta_{kpm}(t) X_p(t) d t+\sum_{j=1}^J w_{kj}^{(1)} z_j+b_k^{(1)}\right) \\
& =g\left(\sum_{p=1}^P \sum_{m=1}^{M_P} c_{kpm} \int_{\mathcal{T}} \zeta_{kpm}(t) X_p(t) d t+\sum_{j=1}^J w_{kj}^{(1)} z_j+b_k^{(1)}\right). 
\end{aligned}
\end{equation}
In Equation \eqref{eqn: FNN general first layer}, the integral can be approximated by using any numerical integration method, e.g., the Simpson's rule \citep{suli2003introduction}. To compute the integral, the functional covariate $X_p(t)$ can be replaced by $\hat{X}_p(t)$ as in Equation \eqref{eqn: expansion approximation}.
The basis coefficients $c_{kpm}$ are differently initialized for each functional weight $\gamma_{kp}(t)$ using the Xavier uniform distribution \citep{glorot2010understanding}, and then will be updated as the FNN learns together with the weights of the other non-functional layers of the FNN.
However, any other weight initialization methods can be used, e.g., see \cite{rastrigin1963convergence, kim1991weight, he2015delving}.
To train and assess the generalization performance of the FNN, the mean squared error (MSE) $\sum_{i=1}^N\left(y_{i}-\hat{y}_{i}\right)^2$ is used as the loss function, where $y_{i}, i=1, \ldots, N$, is the true scalar response and $\hat{y}_{i}$ the FNN fitted output.
The FNN can be trained with the usual backpropagation algorithm \citep{rumelhart1985learning} and the Adam optimizer \citep{kingma2014adam}.  
It is worth emphasizing that the FNN often requires fewer parameters compared to standard NNs (e.g., MLP, CNN, RNN) that directly process raw data. 
For instance, let us consider a single functional covariate, measured $C$ times along its domain.
The first hidden layer of a traditional NN has a number of parameters equal to $(C + 1) \times n^{(1)}$, whereas the FNN requires $(M_1 + 1) \times n^{(1)}$, where $M_1$ is typically lower than $C$ to avoid functional weight to overfitting.

Before training a NN there are many parameters to be specified, sometimes referred to as hyperparameters.
Tuning the values of the hyperparameters plays a critical role in the generalization of the NN model.
Typical hyperparameters are the learning rate, which determines the step size in the optimization process and significantly affects the convergence speed and the possibility of getting stuck in local minima or overshooting global optima; the batch size, which determines the number of samples processed before updating the model's weights, affecting the trade-off between computational efficiency and generalization accuracy;  the activation function, which introduces non-linearity feature and may affect the NN convergence and ability to handle vanishing or exploding gradients \citep{rumelhart1986learning}; the NN architecture, which includes the number of layers and the number of neurons in each layer and significantly influence the ability of the NN to capture complex patterns in the data.
The number and the type of the basis functions, which approximate each functional weight, can be considered hyperparameters as well.
In order to optimize the generalization performance of the FNN, we use a hyperparameter tuning approach combining 5-fold cross-validation \citep{hastie2009elements} with a grid search, which explores all possible combinations of hyperparameters within a grid containing a wide range of values for each hyperparameter of interest.
By combining 5-fold cross-validation and grid search, we ensure comprehensive coverage of the hyperparameter space and identify the optimal parameter values that yield the smallest 5-fold cross-validated MSE, defined as $\sum_{b=1}^5 \sum_{i \in S_b} \left(\hat{y}_i-y_i\right)^2 / N$, where $S_b$ is the $b$th set of observations in the held-out fold, and $\hat{y}_i$, is the predicted value for $y_i$ by the FNN trained on the rest of the $B-1$ folds.
In the subsequent analysis, we tune all the FNN hyperparameters to some degree and then use the early stop strategy \citep{keskar2017improving} to prevent overfitting and improve the generalization performance of the final model. Instead of training the model for a fixed number of epochs, the early stop strategy consists of monitoring the MSE on the \textit{validation set}, which is a separate subset of unseen observations, and stops the training process when the latter increases.

The problem of interpretability in NNs has become a significant concern in the field of DL. While NNs have demonstrated remarkable performance across various tasks, their inherent black-box nature has hindered their wider adoption and acceptance, particularly in domains where interpretability and transparency are paramount. The black box refers to the opacity of NNs, where the internal decision-making processes are not easily understandable or explainable to humans.  
Researchers and practitioners are striving to develop techniques and methodologies that can shed light on the black box, enabling us to gain deeper insights into how NNs arrive at their predictions or classifications. 
The pursuit of interpretability, whose rigorous definition is still debated in machine learning literature \citep{molnar2020interpretable,lepore2022interpretability, stevens2023explainability}, is crucial for ensuring transparency and enabling domain experts to understand, validate, and trust the decisions made by NNs.
To this aim, FNNs guarantee hidden semantic interpretability that refers to the human ability to understand hidden layers \citep{fan2021interpretability}.
Specifically, the functional weights introduced in Equation \eqref{eqn: FNN general first layer} differ from the traditional weights as they can be easily visualized over the continuum, helping the interpretation of the relationship between the functional covariates and the scalar response while preserving the autocorrelation structure associated with the data.  
These functional weights coincide with those estimated in the linear functional regression model reported in Equation \eqref{eqn: sof model least square}.
If the functional hidden layer has more than one neuron, the average of the estimated functional weights $\hat{\gamma}_p(t)=\sum_{k=1}^{n^{(1)}} \hat{\gamma}_{kp}(t) / n^{(1)}$ for each $P$ functional covariate can be considered. 

\subsection{The functional neural network control chart}\label{sec:The monitoring strategy}

We propose a control charting procedure, referred to as FNNCC, to monitor a scalar quality characteristic adjusted by the effect, possibly nonlinear, of one or more functional covariates, which 
relies on the following main steps:
\begin{enumerate}[i]
    \item the relationship between the scalar response and the functional covariates is modeled through an FNN; 
    \item the functional model is estimated through FNN hyperparameter tuning, objective function, and optimization algorithm definition \citep[p.~294-316]{goodfellow2016deep};
    \item the monitoring strategy of residuals $e_i = y_i - \hat{y}_i$, $i=1,2, \ldots, n$ obtained from the FNN is defined. Residuals act as the scalar quality characteristic of interest to indirectly monitor the stability of the functional relationship between the scalar response $y_i$ and the multivariate functional covariates $X_{ip}$. 
    For conciseness of notation, we will hereinafter denote by $\hat{y}_i$  the fitted value of $y_i$, even when an objective function different from MSE (as in Equation \eqref{eqn: sof prediction}) is used.
\end{enumerate}

\begin{figure}
\centering
\includegraphics[scale=0.50]{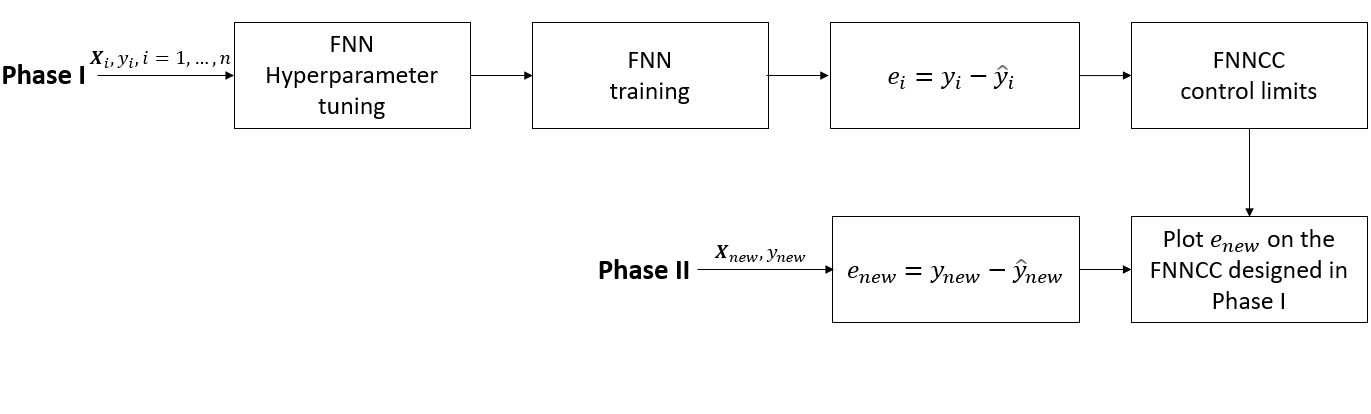}
\caption{Outline of the FNNCC approach.}
\label{fig: FNNCC_scheme}
\end{figure}

A current observation of the scalar response variable $y_i$, given the corresponding functional covariate vector $\mathbf{X}_i(t) = (X_{i1}, \ldots, X_{iP})$, is monitored by using the FRCC approach, that is the FNNCC results in a univariate control chart based on the FNN residual $e = y - \hat{y}$. 
This paper focuses on the prospective monitoring of the residuals, referred to as Phase II.
That is, a data set of observations representative of the IC process performance, referred to as \textit{Phase I sample} or \textit{reference data set}, is assumed to be available.
It is worth noting that the retrospective monitoring, referred to as Phase I is crucial to check the stability of historical functional data and to obtain accurate estimates of the unknown parameters used for Phase II monitoring \citep{zhang2015phase}.
The reference data set is randomly split into three non-overlapping sets, referred to as \textit{training}, \textit{validation}, and \textit{tuning} sets.
The first is used to design and train the FNN; the second to implement the early stopping strategy \citep{goodfellow2016deep}, which allows stopping the training when the performance does not improve and thus, to prevent overfitting and improve the generalization of the NN; the third is used to estimate the upper and lower control limits (CLs), respectively, as  $\alpha/2$ and $(1-\alpha/2)$ empirical quantiles of the sampling distribution of the FNN residuals estimated from the tuning set, being $\alpha$ the type-I error rate \citep{qiubook}.
In Phase II, the residual of a new observation ($\mathbf{X}_{new}, y_{new}$) is calculated as
\begin{equation}
\label{eqn:residual new observation}
e_{new} = y_{new} - \hat{y}_{new},
\end{equation}
where $\hat{y}_{new}$ is the value fitted by the FNN model identified in Phase I, according to the FNN model estimated with choices made in steps (i) and (ii).
An alarm is issued if $e_{new}$ is larger than UCL or lower than LCL.
The implementation of the FNNCC is outlined in  Figure \ref{fig: FNNCC_scheme}.

If the FNNCC issues an alarm, a change in the relationship between the response and the covariates may have occurred. This could be due to changes in the regression coefficients associated with one or more covariates, or potential causes outside the set of covariates included in the model may be investigated \citep{shu2004run}.

\section{Simulation study}\label{sec: Simulation study}

The overall performance of the proposed FNNCC is compared with FRCC and with a univariate Shewhart control chart, referred to as SCC, which monitors the scalar response without considering any information on the functional covariates, in terms of $ARL_1$ at a given $ARL_0$ \citep{qiubook} by means of an extensive Monte Carlo simulation.
Without loss of generality, the compact domain $\mathcal{T}$ is assumed as $[0,1]$, and the number of covariates $P$ is set equal to 1. 
Then, the scalar response observations are generated from an IC process as in the following scenarios
\begin{eqnarray}
\text{Scenario A} \quad y^*&=&  G(\mathbf{X}_i) + \epsilon^* = \alpha+\int_{\mathcal{T}} \beta(t) x(t) \mathrm{d} t+\epsilon^*\\
\text{Scenario B} \quad y^*&=&  G(\mathbf{X}_i) + \epsilon^* = \exp \left(\alpha+\int_{\mathcal{T}} \beta(t) x(t) \mathrm{d} t\right)+\epsilon^* \\
\text{Scenario C} \quad y^*&=&  G(\mathbf{X}_i) + \epsilon^* = \Bigg | \alpha+\int_{\mathcal{T}} \beta(t) x(t) \mathrm{d} t \Bigg | +\epsilon^* \\
\text{Scenario D} \quad y^*&=&  G(\mathbf{X}_i) + \epsilon^* = \log \left(\left|\alpha+\int_{\mathcal{T}} \beta(t) x(t) \mathrm{d} t\right| + u \right)+\epsilon^*\\
\text{Scenario E} \quad y^*&=&  G(\mathbf{X}_i) + \epsilon^* =  \left(\alpha+\int_{\mathcal{T}} \beta(t) x(t) \mathrm{d} t\right)^2+\epsilon^*, \\
\end{eqnarray}
where the noise $\epsilon^*$ is sampled from the Gaussian distribution $N(0, 0.1)$.
where in each scenario, apart from Scenario A, $G(\cdot)$, defined in Equation \eqref{regression model}, specializes to a nonlinear SOF mapping, and $\epsilon_i^*$  is the error term.
It is worth noting that in Scenario D, $u$ is an arbitrary positive constant, which is set equal to 2 to avoid numerical problems due to small values.
$y_i$ and $\alpha+\int_{\mathcal{T}} \beta(t) x_i(t) \mathrm{d} t$ are generated using the functions \texttt{simulate\_mfd()}, which is called by the wrapper function \texttt{sim\_funcharts()} from the \texttt{funcharts} package \citep{capezza2022funcharts}.
The mean and the variance function $\mu^{X}(t)$ and $v^{X}(t)$ of the functional covariates are generated according to the following model
\begin{equation}
\label{eqn: reference model}
f(z)=P(z)+r \sum_{i=1}^I h_i\left(z ; m_i, s_i\right), \quad z \in(0,1),
\end{equation}
with
\begin{equation}
\label{eqn: pz reference model}
   P(z)=a z^2+b z+c,
\end{equation}
where $a, b$ and $c$ are real numbers, and the terms $h_i\left(z ; m_i, s_i\right)$ are normal probability density functions with mean $m_i$ and standard deviation $s_i$.
The functional covariate ${X}(t)$ is characterized by the Bessel \citep{abramowitz1964handbook} correlation function and is evaluated at 150  equally spaced discrete points of the functional domain $\mathcal{T} = [0,1]$.
As the simulated functional covariate data are observed at noisy discrete values, each functional observation is obtained by Equation \eqref{eqn: expansion approximation} with $M=30$ cubic B-splines estimated through the spline smoothing approach. 
Then, the scalar response is generated through a SOF linear model so that the determination coefficient $R^2$ defined by \cite{yao2005functional}, which measures the proportion of the variance in the response variable explained by the functional covariate in the model, is set equal to 0.97.
The mean and the variance of $y$ are set to  $\mu^y = 0$ and $v^y=1$, respectively.  

The performance of the proposed control chart is studied under a shift in $\mu^y$ only, and both
in the functional covariate mean $\mu^X(t)$ and $\mu^y$.
In the latter case, to study the unwanted effect of the shift in $\mu^X(t)$ on the FNNCC performance, a translation of the profile pattern is generated by using the model defined in Equation \eqref{eqn: reference model} with $P(z)$ defined as 
\begin{equation}
\label{eqn: translation pattern}
P(z)=a z^2+b z+(c+\delta), \quad z \in(0,1), 
\end{equation}
where $\delta$ is a real number defining the translation magnitude and is set to be equal to 0.5. 
In both scenarios, the mean shift in the scalar response is obtained by adding, to the simulated values, a fixed quantity that defines the mean shift size.
That is, $\mu_y$ is shifted by as much as $\Delta \mu_y = \{0.5,1,1.5,2 \}s^{y^*}$, where $s^{y^*} = \sqrt{(v^{y^*})^2}$ is the standard deviation of the transformed scalar response. 

For each simulated scenario, a set of 4000, 1000, and 10000 IC scalar responses and functional covariates are randomly generated to form the training, validation, and tuning sets, respectively, for the reason discussed in Section \ref{sec:The monitoring strategy}.
To evaluate the $ARL_1$, an additional set of 20000 OC patterns is randomly generated.    
$ARL_0$ is set to 20, which corresponds to $\alpha = 0.05$. 
In all scenarios, based on 5-fold cross-validation with grid search, we use in step (ii) a two-layer FNN with 8 neurons and with ReLu and linear activation functions. Each functional coefficient is expressed as a linear combination of 5 cubic B-spline (see Equation \eqref{eqn: beta expansion}). 
Then, the FNN is trained in step (ii) using the backpropagation algorithm with the Adam optimizer to minimize the MSE chosen as the objective function.
For the FRCC, regression coefficients are estimated using the training set, while the tuning set is used to compute the CLs as the $\alpha/2$ and $1-\alpha/2$ of empirical quantiles of the scalar response.
\begin{figure}
\centering
\begin{minipage}{0.31\textwidth}
\includegraphics[width=\linewidth]{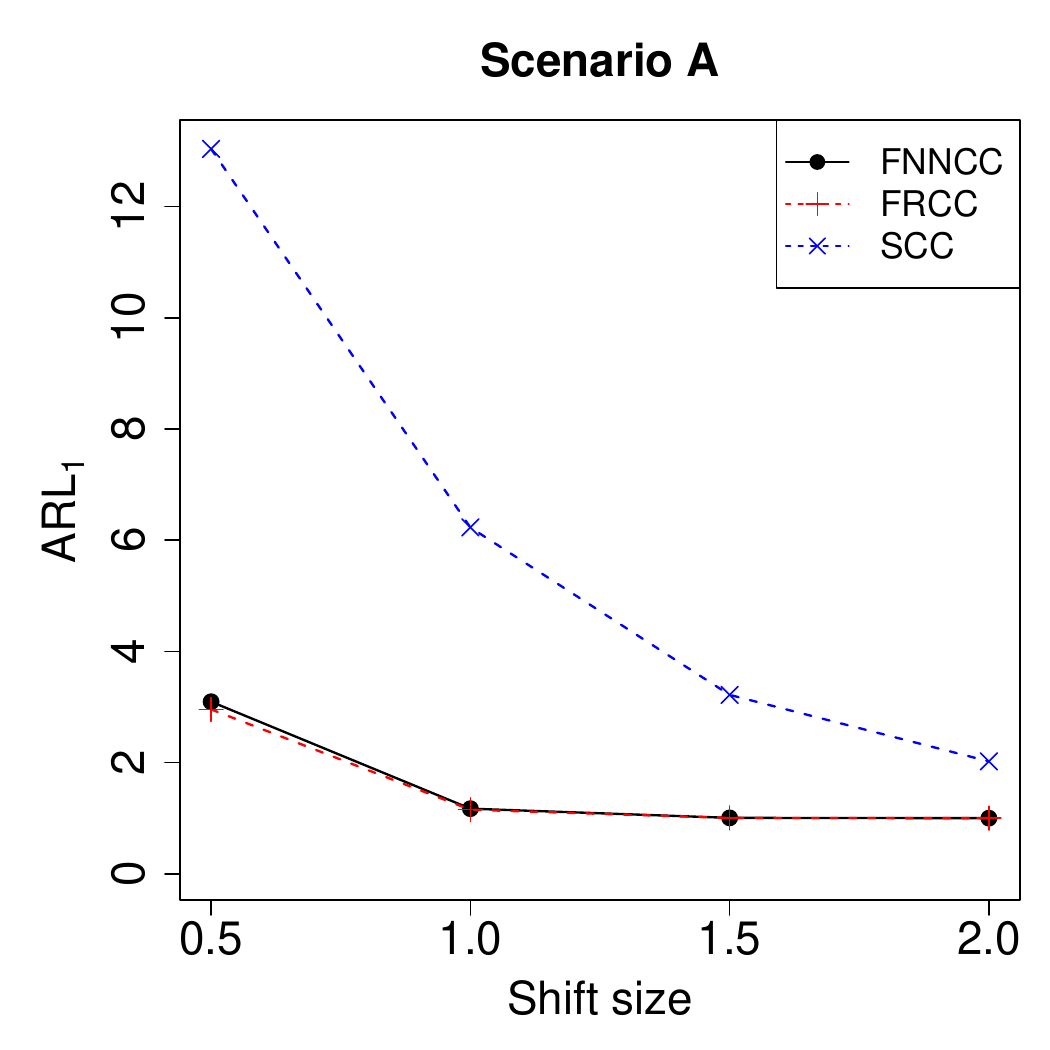}
\label{fig: linear FNNCC vs sofRCC and SCC}
\end{minipage}
\hspace*{\fill}
\begin{minipage}{0.31\textwidth}
\includegraphics[width=\linewidth]{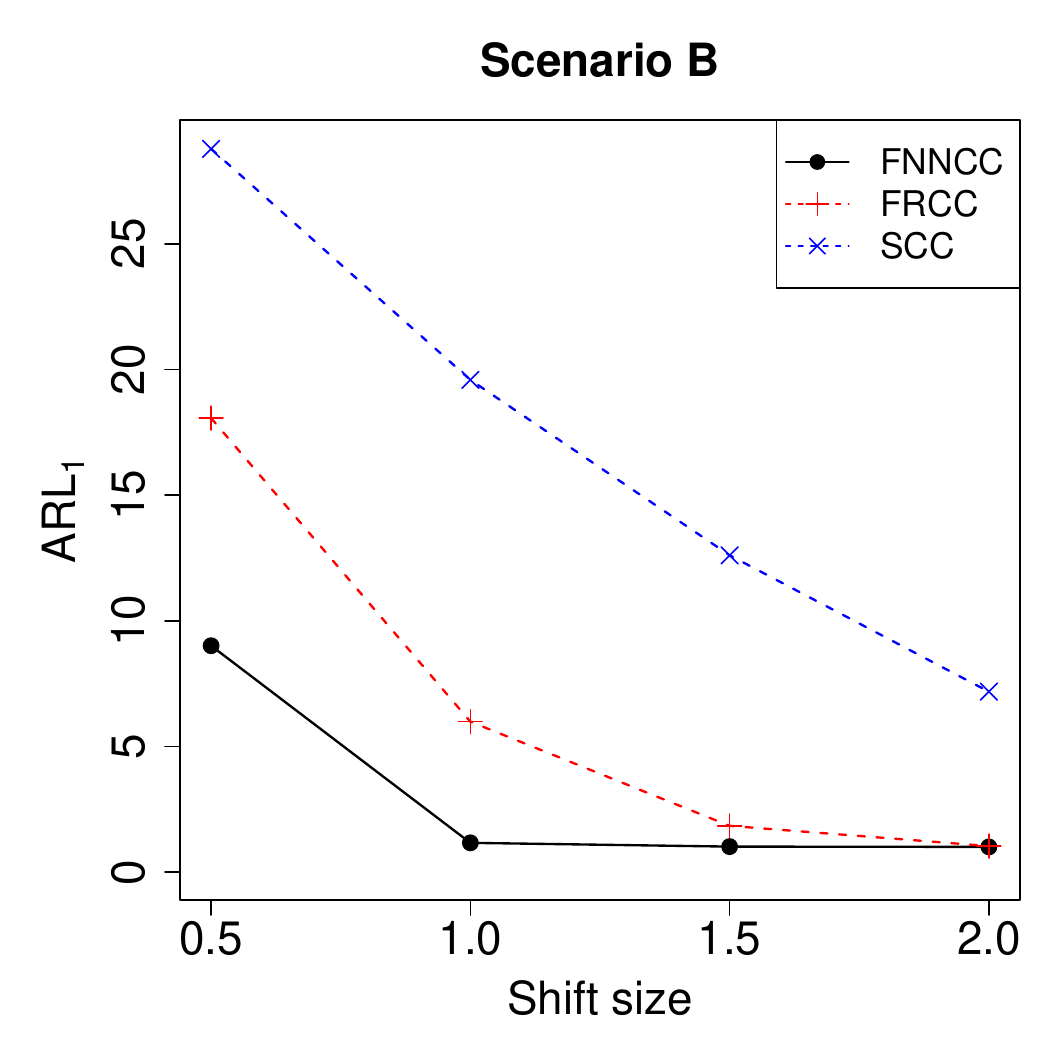}
\label{fig: exponential FNNCC vs sofRCC and SCC}
\end{minipage}
\hspace*{\fill}
\begin{minipage}{0.31\textwidth}
\includegraphics[width=\linewidth]{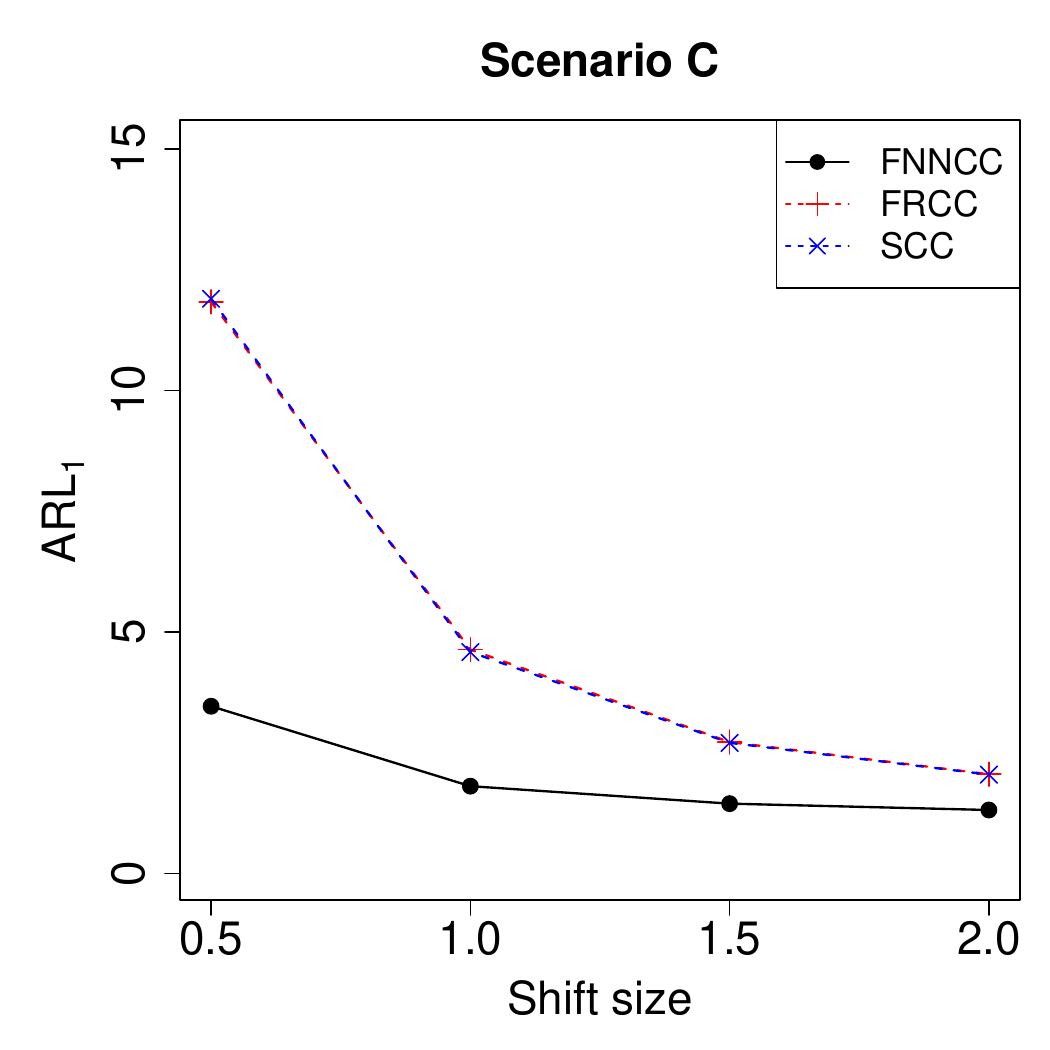}
\label{fig: abs FNNCC vs sofRCC and SCC}
\end{minipage}
\bigskip
\begin{minipage}{0.31\textwidth}
\includegraphics[width=\linewidth]{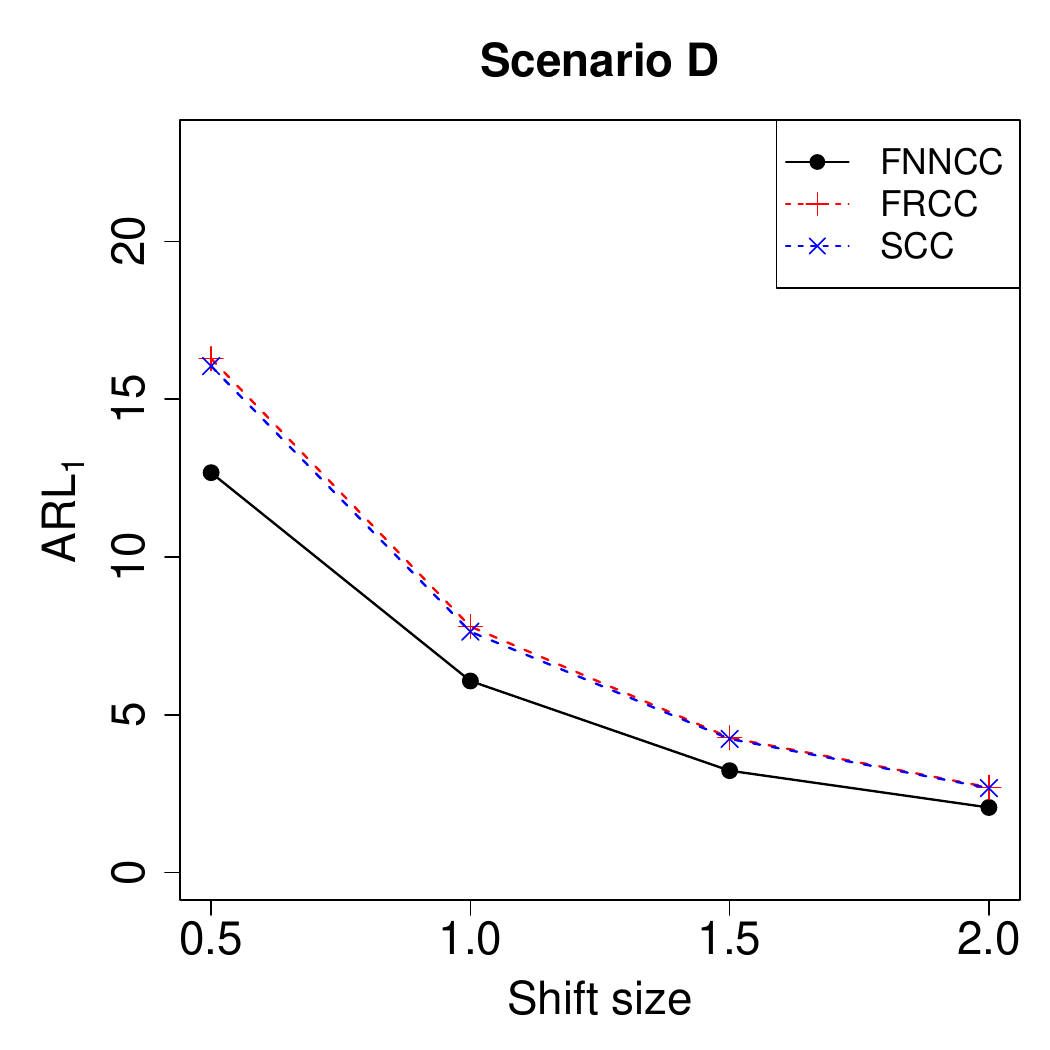}
\label{fig: log FNNCC vs sofRCC and SCC}
\end{minipage}
\hspace{3mm} 
\begin{minipage}{0.31\textwidth}
\includegraphics[width=\linewidth]{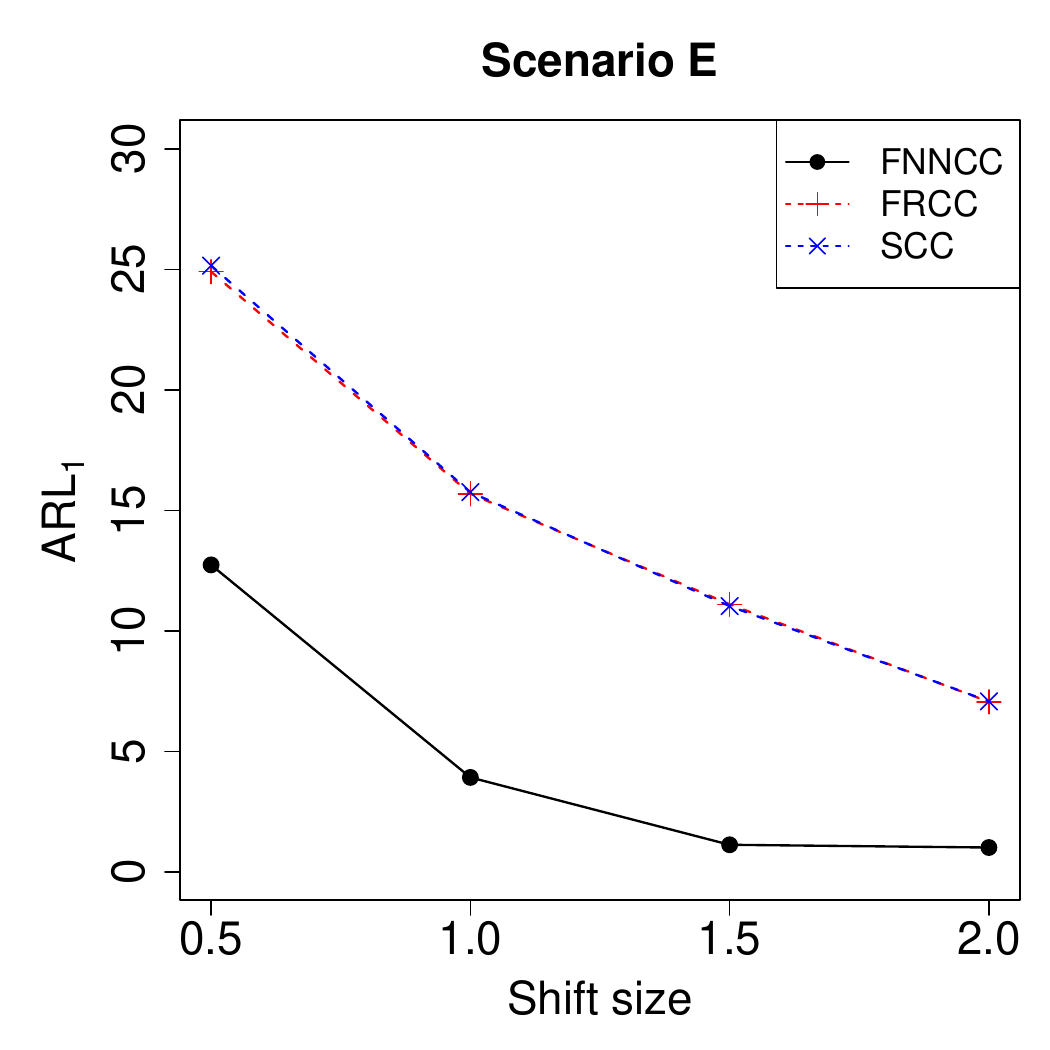}
\label{fig: square FNNCC vs sofRCC and SCC}
\end{minipage}
\caption{Estimated $ARL_1$ achieved by FNNCC, FRCC, and SCC for each simulated scenario, as a function of the mean shift size of the scalar response $\Delta \mu_y = {0.5,1,1.5,2s^{y^*}}$.}
\label{fig: FNNCC vs sofRCC and SCC}
\end{figure}
%
\begin{figure}[!h]
\centering
\begin{minipage}{0.31\textwidth}
\includegraphics[width=\linewidth]{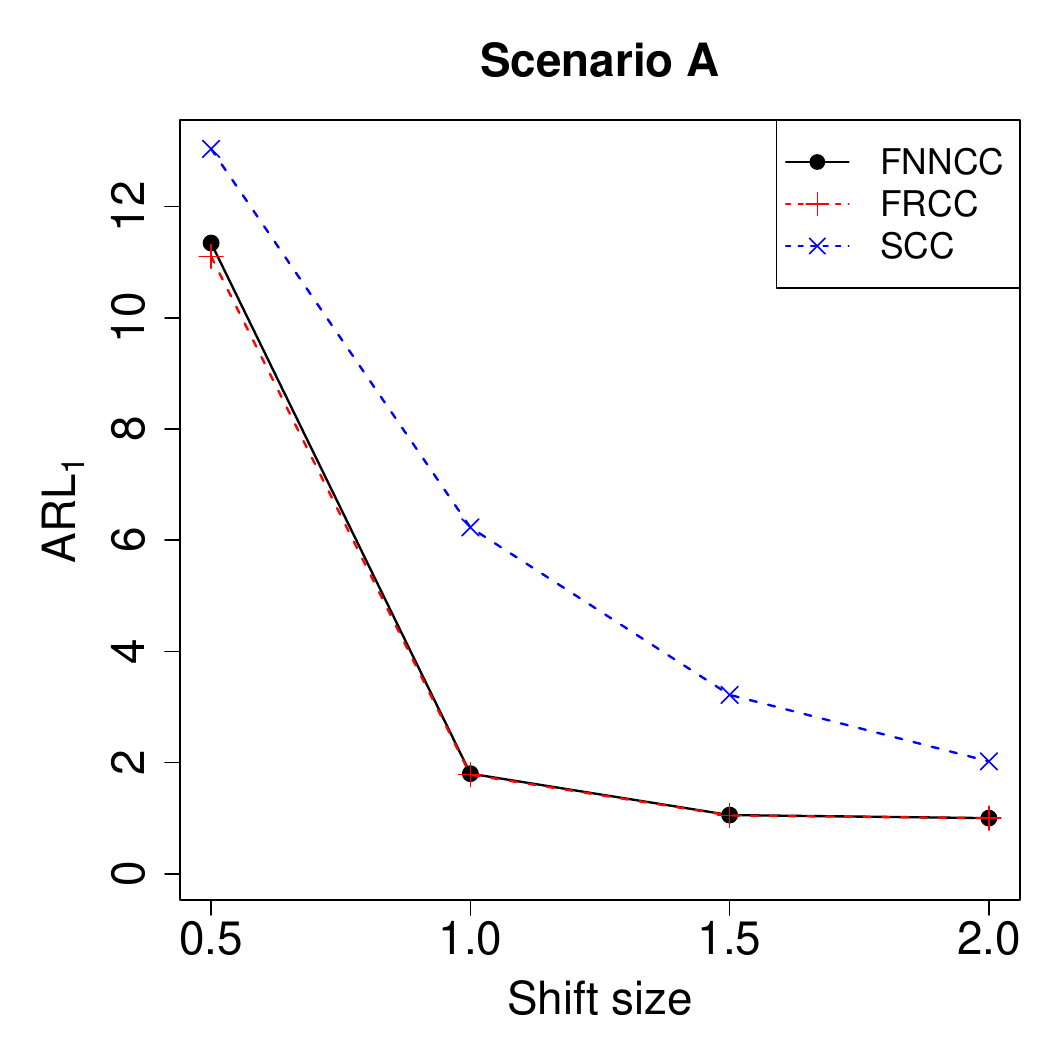}
\label{fig: linear FNNCC vs sofRCC and SCC cov shift}
\end{minipage}
\hspace*{\fill}
\begin{minipage}{0.31\textwidth}
\includegraphics[width=\linewidth]{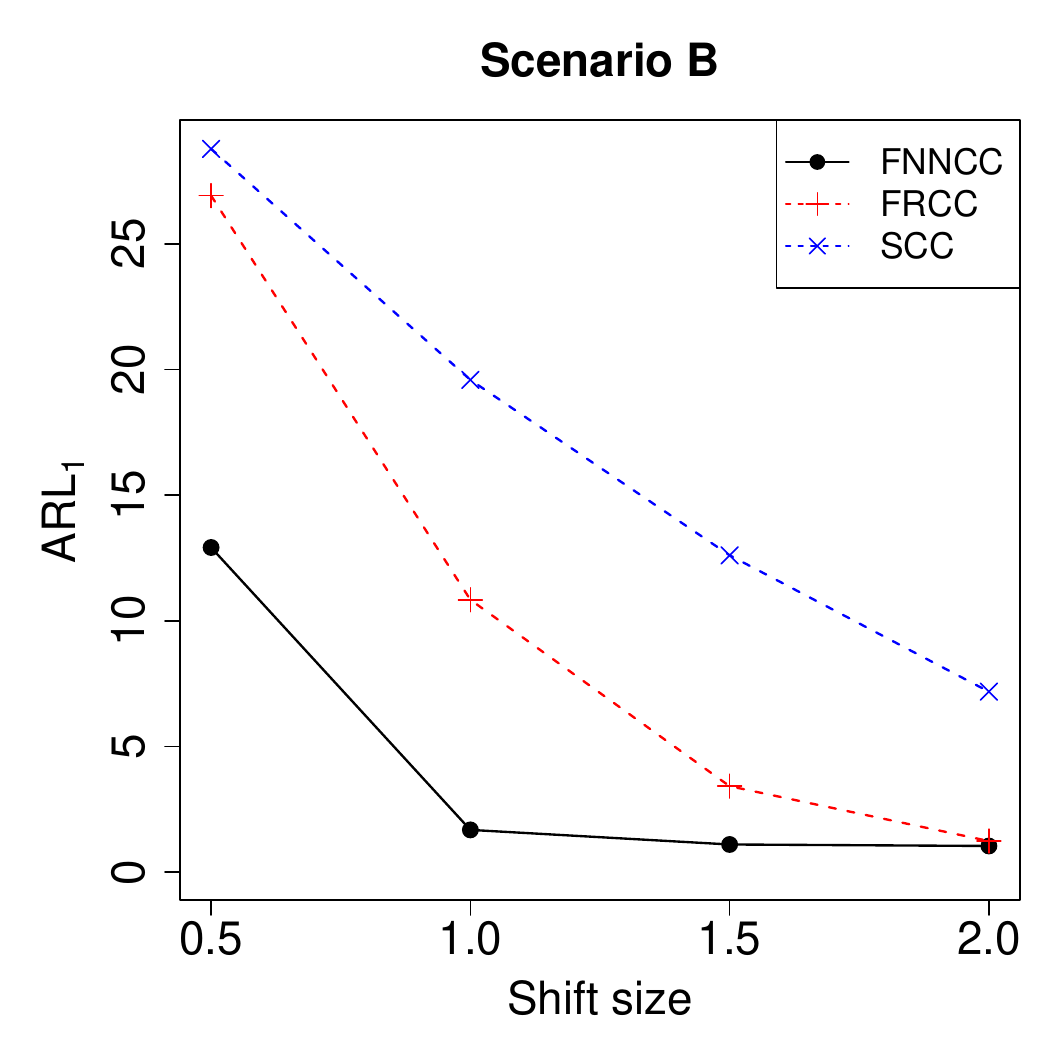}
\label{fig: exponential FNNCC vs sofRCC and SCC cov shift}
\end{minipage}
\hspace*{\fill}
\begin{minipage}{0.31\textwidth}
\includegraphics[width=\linewidth]{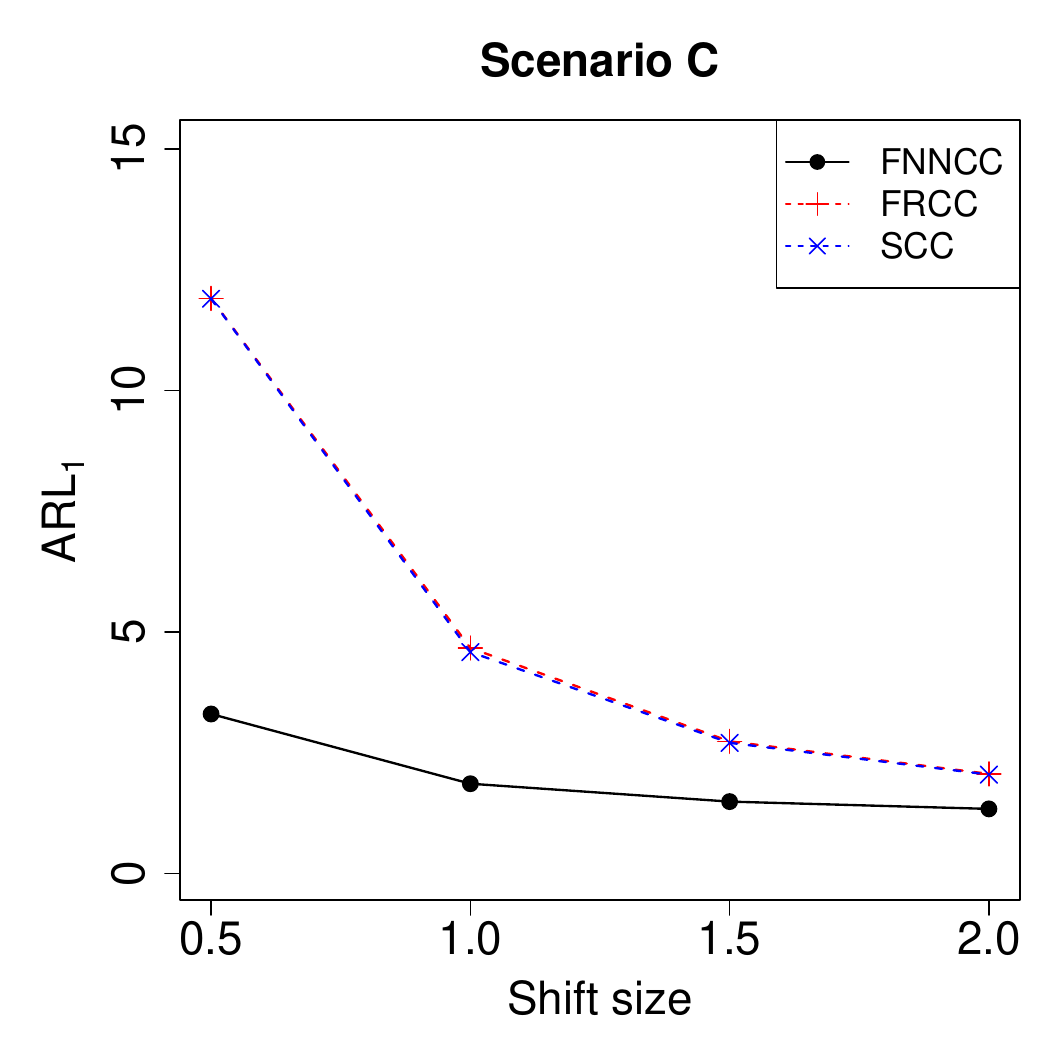}
\label{fig: abs FNNCC vs sofRCC and SCC cov shift}
\end{minipage}
\bigskip
\begin{minipage}{0.31\textwidth}
\includegraphics[width=\linewidth]{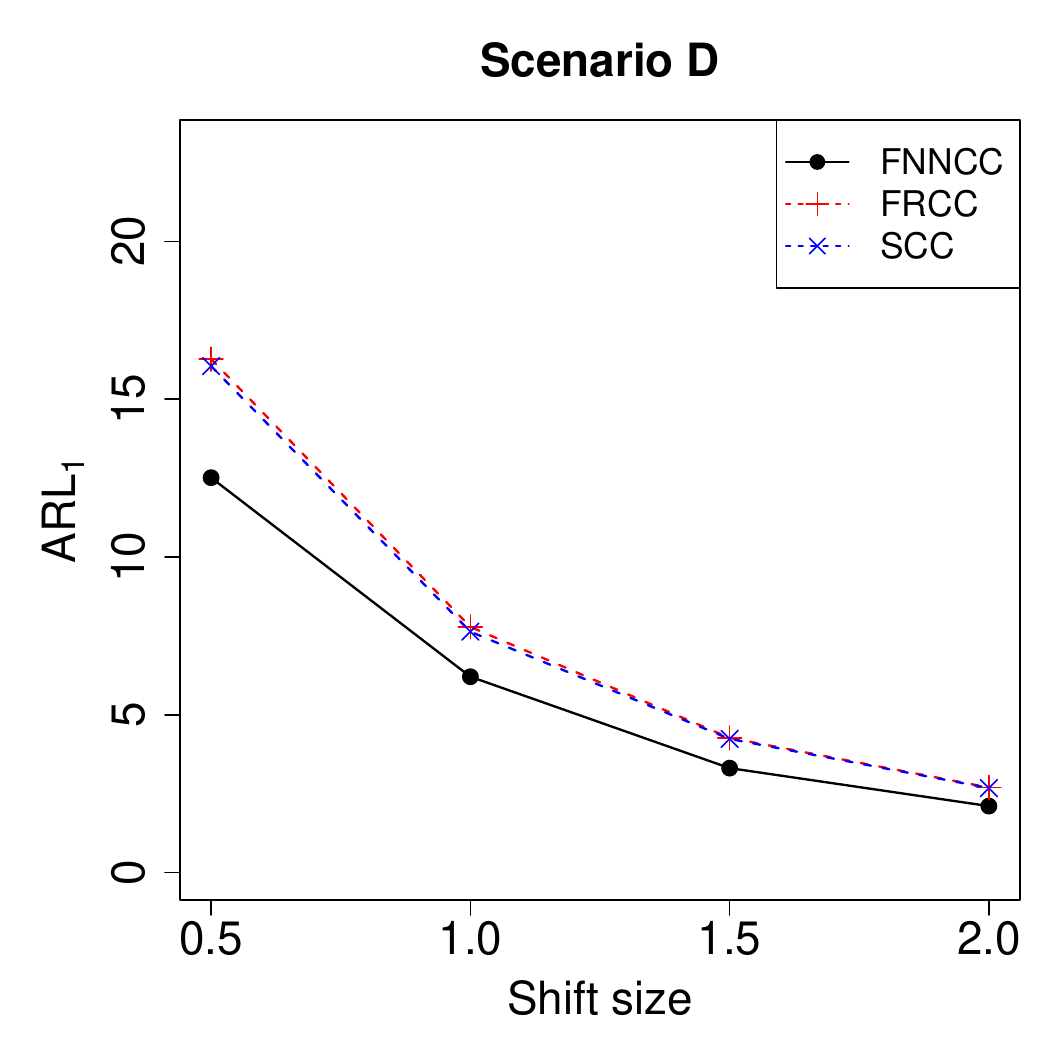}
\label{fig: log FNNCC vs sofRCC and SCC cov shift}
\end{minipage}
\hspace{3mm} 
\begin{minipage}{0.31\textwidth}
\includegraphics[width=\linewidth]{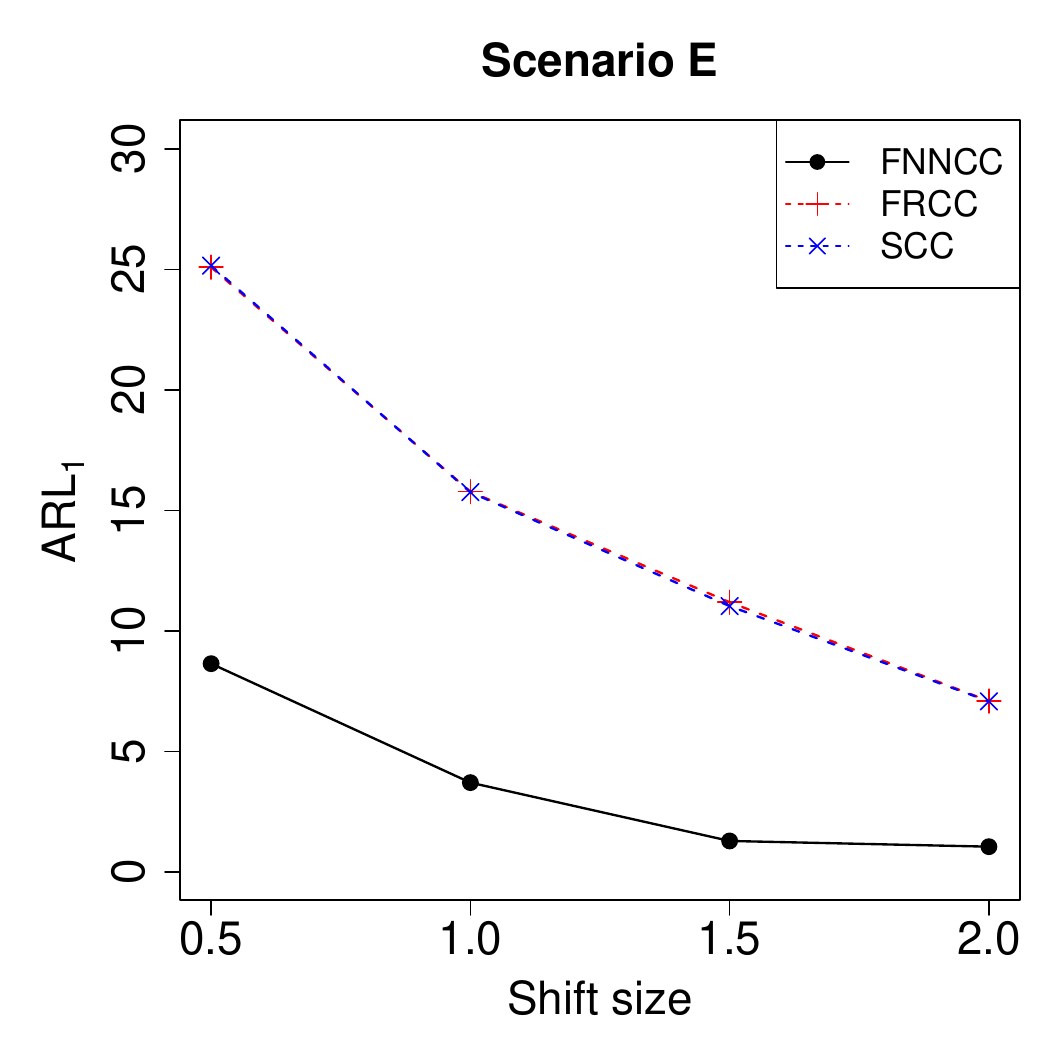}
\label{fig: square FNNCC vs sofRCC and SCC cov shift}
\end{minipage}
\caption{Estimated $ARL_1$ achieved by FNNCC, FRCC, and SCC for each simulated scenario, as a function of the mean shift size of the scalar response $\Delta \mu_y =  \{0.5,1,1.5,2 \} s^{y^*}$ when the functional covariate is subject to a mean shift}
\label{fig: FNNCC vs sofRCC and SCC cov shift}
\end{figure}

Figures \ref{fig: FNNCC vs sofRCC and SCC} graphically represent the $ARL_1$ performance achieved by FNNCC, FRCC, and SCC for each simulated scenario, as a function of the mean shift size of the scalar response $\Delta \mu_y = \{0.5,1,1.5,2 \} s^{y^*}$, when no covariate mean shift is considered. 
As we would expect, the SCC shows the worst performance for all the considered scenarios and shifts as it is not capable of adjusting the monitoring of the scalar response by the effect of the additional information provided by the functional covariates.
In Scenario A, the FRCC performs comparably to the FNNCC as the linear functional regression model is able to capture the true linear relationship between the scalar response and the functional covariate. 
In all other scenarios, the FNNCC outperforms the FRCC for all the considered shifts.
This confirms the SOF regression model is not capable of modeling the true nonlinear relationship and, thus, the FRCC performance is the same as the SCC.
For example, for Scenario C, the $ARL_1$ of the FNN, FRCC, and SCC are 3.46, 11.83, and 11.89, respectively, at $\Delta \mu_y = 0.5s^{y^*}$.  
The gain in efficiency decreases as the shift size increases.
As an example, when the relation between the response and the functional covariate is modeled by an exponential function (Scenario B), the $ARL_1$ of the FNNCC and of the FRCC are 9.01 and 18.06 at a small shift size $\Delta \mu_y = 0.5s^{y^*}$, respectively, whereas for a higher shift size, say $\Delta \mu_y = 1.5s^{y^*}$,  their performance decrease to 1.02 and 1.84, respectively. 
When the functional covariate is subject to a mean shift, Figure \ref{fig: FNNCC vs sofRCC and SCC cov shift} displays the estimated $ARL_1$  of the three competing control charting schemes for all the simulated scenarios, as a function of the size of the response mean shift.
This figure points out that shifts in the covariate mean function may impact the $ARL_1$ of the FNNCC and FRCC. 
It is trivial to note that the SCC performance is unaffected by a change in $\mu^x(t)$.
Simulation results, displayed in Figure \ref{fig: FNNCC vs sofRCC and SCC cov shift}, show that the FNNCC still results more sensitive than the FRCC in detecting OC condition of the scalar response and that the $ARL_1$ performance of the FNNC and FRCC generally increases (or at least remains the same) in the presence of a covariate mean shift. 
These results are consistent with \cite{centofanti2021functional, shu2004run}.

The simulation study clearly highlights the superiority of the proposed method in dealing with nonlinearity. As the true nature of the relation
between the scalar response and the functional covariates is never known in real-world processes, FNNCC has proven to be a more flexible strategy to be recommended in situations where the influence of the functional covariates is not necessarily linear.

Moreover, by means of an extensive Monte Carlo simulation, the FNNCC is compared with other two NN-based monitoring strategies implemented by using an MLP in place of the FNN in step (i) of the monitoring strategy described in Section \ref{sec:The monitoring strategy}.
While both methods exhibit similar performance, the FNNCC is the preferred choice due to its interpretability. The functional coefficients of the FNN enable visualizing the relationship between functional covariates and the scalar response. In contrast, the weights and intercepts of MLPs are challenging to interpret, and the existing literature on interpreting conventional NNs is primarily limited to computer vision applications.
The details of the numerical analysis are reported in Appendix A.

\section{Case study: monitoring of HVAC systems on modern passenger trains} \label{sec: Real-Case Study}

The case study mentioned in the Introduction is presented to demonstrate the applicability of the proposed control chart in real situations. 
In recent years, European regulations have been established to set operational standards for the thermal comfort of passenger rail coaches. These standards, such as \cite{en200614750}, were developed to meet the different operating requirements of rail vehicles and to ensure a high-quality air environment for passengers.
In view of those standards, railway companies are increasingly installing sensing systems to collect data from onboard HVAC systems.

HVAC systems regulate the indoor temperature of each train's coach through a combination of ventilation, heating, and cooling operations. 
Ventilation is the process of replacing or exchanging indoor air with outdoor air to remove harmful particles like dust, smoke, and bacteria.
Heating and cooling, on the other hand, provide warmth or cold inside the coaches. 
An HVAC system has three main components: the compressor, the condenser, and the thermal expansion valve. 
The compressor moves the refrigerant gas to the condenser, where the gas changes into a liquid. 
Then, the liquid refrigerant moves through the evaporator section, where it evaporates into a cold gas, absorbing heat from the surrounding air and cooling down the coach interior.
Eventually, the thermal expansion valve converts the cold gas back into a liquid, and the process repeats multiple times until the indoor temperature reaches the desired level. 

The five passenger trains object of this study have six coaches, which are equipped with a dedicated HVAC system, A central unit is installed to control the heating and cooling modes of each HVAC system based on temperature sensors that measure changes in the outdoor ($T_\text{out}$) and indoor ($T_\text{in}$) temperature signals that are streamed to maintenance engineers for monitoring purposes and potentially improving reliability and maintenance programs.
At each time instant, on each coach, the HVAC system activates until  $T_\text{in}$ matches the target temperature ($T_\text{set}$), which is automatically and independently set as a function of $T_\text{out}$, to comply with the current regulation on passengers' comfort.
The raw measurements of these temperatures are contained in the \texttt{HVAC} data set where, for confidentiality reasons, the train names, acquisition year, and routes are omitted. Trains and coaches are then identified with a number from 1 to 5 and 1 to 6, respectively. 
This case study focuses on one specific route where temperature measurements are acquired at a regular grid of points equally spaced by 30 seconds.
Railway engineers confirmed that the 30 HVAC systems can be assumed to be equal and working under the same operating conditions (e.g., outdoor temperature, and passenger load) on the same route. 
Thus, the coach effect is not considered hereinafter. 
Raw measurements are then grouped to form raw profiles referred to different voyages, which are identified by a unique voyage number (VN). 
Based on experts’ opinion, exceptional voyages that do not represent normal operating conditions have been removed to define the Phase I sample that is thus formed by 1853 voyages and, according to Section \ref{sec:The monitoring strategy},  is randomly split into 740, 186, and 927 voyages to form training, validation, and tuning sets, respectively.

Temperature profiles are re-mapped as a function of the fraction of the total distance traveled by train at each voyage, and, to avoid the modeling of transitional regimes, due to the HVAC restart at the beginning of each voyage, the first $25\%$ of the traveled distance of each profile is discarded.
This operation can be regarded as a landmark registration \citep[p.~129-132]{ramsaybook} of the functional data set from the function-specific temporal domain to the common domain $\mathcal{T}$, which, without loss of generality, is set as $[0,1]$.
To get smooth profiles, i.e., functional data observations, we choose a B-spline basis system with 70 basis functions and equally spaced knots estimated by solving a regularization problem with a roughness penalty on the integrated-squared second derivative and smoothing parameter chosen through a generalized cross-validation \citep[p.~97-
99]{ramsaybook}.

The scalar quality characteristic of interest is the root mean square (RMS) of the difference between measurements $T_\text{in} and T_\text{set}$ acquired during each train voyage, referred to as \texttt{DevTemp}.
That is, the scalar response, \texttt{DevTemp}, is computed once the voyage is ended as engineers cannot perform maintenance operations until the train has finished its voyage or reached the terminal station. 
\texttt{DevTemp} is assumed to be influenced by $T_\text{out}$ and the $T_\text{set}$ derivative (with respect to the fraction of the total distance travel), hereinafter denoted by $\dot{T}_\text{set}$.
The former is included in the model to the extent of accounting for the thermal load under which the system works, whereas the latter, based on experts' opinion, is able to account for the thermal inertia of the process, which is the time the HVAC system needs to allow the $T_\text{in}$ to match $T_\text{set}$.
$T_\text{set}$ may indeed vary as a function of $T_\text{out}$, which, in turn, may be subject to rapid changes, e.g., in under and over-ground route segments.

\begin{figure}
\label{fig: training profiles}
\centering
\includegraphics[scale=0.95]{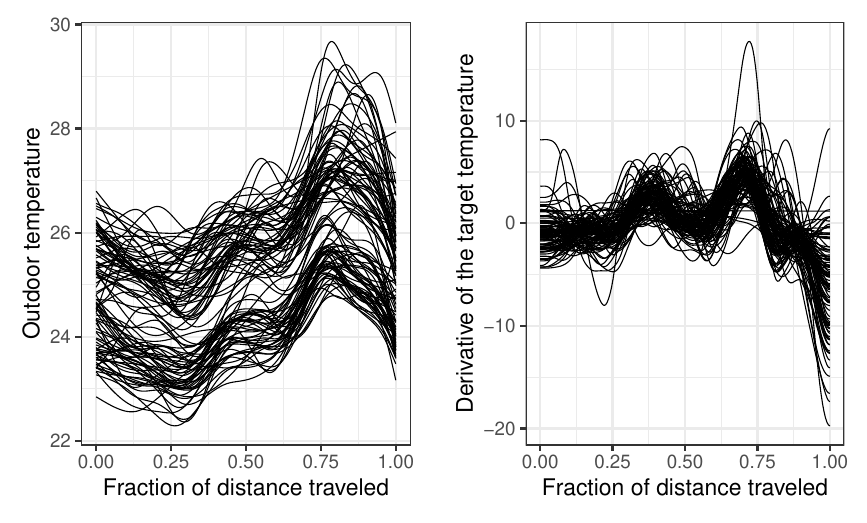}
\caption{100 random samples of the two functional profiles from the HVAC training set.} 
\end{figure}

\begin{figure}
\centering
\includegraphics[scale=0.8]{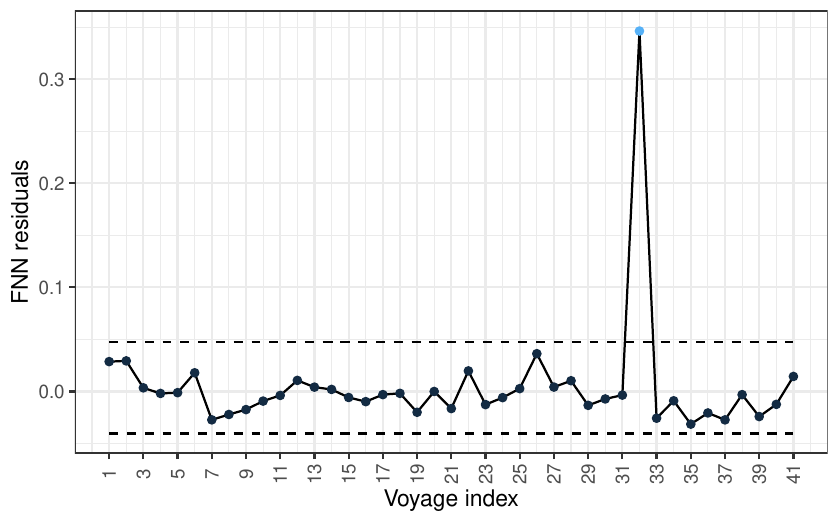}
\caption{Phase II FNNCC. Each point corresponds to a voyage and the values of the corresponding residual are reported. Horizontal dashed lines are the CLs and the point above the CLs denotes the OC observation.} 
\label{fig: FNNCC}
\end{figure}

For illustrative purposes, Figure \ref{fig: training profiles} displays a random slice of 100 observations from the training sample of the $T_\text{out}$ and $T_\text{set}$ derivative profiles.  
The FNN is trained using a backpropagation algorithm with Adam optimizer and early stopping, with the same set of hyperparameters discussed in the simulation study in Section \ref{sec: Simulation study}. 
Specifically, we first use the tuning data set to build the FNNCC and estimate the CLs with a type-I error rate $\alpha = 0.05$.
41 additional profiles from \texttt{coach 4} of \texttt{train 5}, which are known to contain a fault due to a diagnosed failure in one of the two HVAC compressors, are used as Phase II observations.
Figure \ref{fig: FNNCC} shows the Phase II monitoring of the aforementioned voyages.
The $x$-axis label is the VN, while the FNN residuals, defined in Section \eqref{sec:The monitoring strategy}, are reported on the $y$-axis.
Voyage 32 shows an overly large value of the monitoring statistics, thus correctly signaling the OC state of the corresponding HVAC system, which was promptly repaired by train maintenance service. 
Indeed, subsequent voyages plot inside the CLs, further demonstrating the proposed control chart to properly track both the IC and OC states in practice.


\section{Conclusions}\label{Conclusions}

A novel profile monitoring charting scheme is proposed in this work and referred to as \textit{functional neural network control chart} (FNNCC). 
Based on a deep learning (DL) model, the FNNCC can adjust the monitoring of a scalar quality characteristic of interest for the nonlinear effect of the functional covariates when are available.
Specifically, the FNNCC relies on functional neural network (FNN), which has recently appeared in the DL literature to allow a neural network to learn possibly nonlinear relationships based on covariates in the form of functional data, aka profiles.  
The residuals obtained from the FNN are elaborated to build a functional regression control chart (FRCC), which also appeared in the statistical literature, although implemented only under the (functional) linearity assumption. The proposed FNNCC is thus the first DL-based profile monitoring scheme that can efficiently exploit additional information on functional covariates, in a possibly nonlinear fashion.

An extensive Monte Carlo simulation is carried out to assess the performance of the proposed FNNCC in identifying mean shifts in the quality characteristic of interest, which is in a scalar form, in the presence or absence of covariate mean shifts.
Then, the FNNCC is compared with the FRCC and the Shewhart control chart (SCC) built on the scalar response. The results show that the FNNCC is far better than the two competitors when the relation between the scalar response and the functional covariate is nonlinear.
Additionally, the FNNCC is compared with two NN-based control charting strategies and, even though they show similar performance, the FNNCC is preferred due to its interpretability of the functional coefficients.
The practical applicability of the proposed control chart is finally illustrated through a case study in the monitoring of heating, ventilation and air conditioning systems installed on board six coach passenger trains, where the favorable performance of the proposed method in properly tracking the IC and OC states of the 
process is shown in practice. 

The integration of neural networks into functional data analysis remains an interesting topic as it efficiently enables nonlinear learning when the covariates are in the form of profiles. 
Future research can be addressed to extend the FNNCC to different and more sophisticated monitoring statistics. 

\section*{Acknowledgements}

The authors are extremely grateful to the Operation Service and Maintenance Product Evolution Department of Hitachi Rail Italy  S.p.A. and, in particular, to engineers Giuseppe Giannini, Vincenzo Criscuolo, and Guido Cesaro for their technological insights in the interpretation of results.

\noindent This work was supported by the MOST – Sustainable Mobility National Research Center and received funding from the European Union Next-GenerationEU (PIANO NAZIONALE DI RIPRESA E RESILIENZA (PNRR) – MISSIONE 4 COMPONENTE 2, INVESTIMENTO 1.4 – D.D. 1033 17/06/2022, CN00000023). This manuscript reflects only the authors’ views and opinions, neither the European Union nor the European Commission can be considered responsible for them.

\noindent The computing resources and the related technical support used for this work have been provided by CRESCO/ENEAGRID High-Performance Computing infrastructure and its staff \citep{ponti2014role}. CRESCO/ENEAGRID High-Performance Computing infrastructure is funded by ENEA, the Italian National Agency for New Technologies, Energy and Sustainable Economic Development, and by Italian and European research programs, see \url{http://www.cresco.enea.it/english} for information.

\appendix
\section{Additional simulation study}
\label{sec: Appendix Additional simulation study}
In the simulation study, the analyses were carried out by comparing the performance of the FNNCC with that of the FRCC and SCC in terms of the $ARL_1$. 
In this Appendix, additional simulations are run for the nonlinear scenarios (Scenario B-E) at each shift type and size discussed in Section \ref{sec: Simulation study}, to compare the proposed control chart with other two NN-based control charting schemes, implemented by using an MLP in place of the FNN at step (i) of the monitoring strategy discussed in Section \ref{sec:The monitoring strategy} to model the nonlinear relationships between the scalar response and the functional covariate.
In particular, in the first control charting strategy, referred to as RawdataMLPCC, the raw discrete values of the functional covariate are directly fed into the MLP \citep{rossi2002functional}, whereas the second competitor, namely BsplineMLPCC, involves a pre-processing step of the functional covariates into a vector of scores from its B-spline expansion \citep{rossi2005representation}. 
The two MLPs, with the same set of the FNN hyperparameters defined in the simulation study in Section \ref{sec: Simulation study}, are trained on the training set using the backpropagation algorithm with the Adam optimizer, and the CLs of the two competing methods are estimated on the tuning set. 
Figures A1-A4 show the estimated $ARL_1$ for the FNNCC, RawdataMLPCC, and BsplineMLPCC for the nonlinear scenarios from B to E, respectively, both in the presence and in the absence of a mean covariate shift.
It is clear from these figures that the RawdataMLPCC shows worse performance than the \textit{functional} counterparts as it is not able to account for the functional nature of the covariate.
FNNCC and BsplineMLPCC achieve similar performance for all OC scenarios.
However, the implementation of the FRCC is to be preferred thanks to its easier interpretability of the functional coefficients.
\begin{figure}[hbt!]
\centering
\begin{minipage}{0.41\textwidth}
\includegraphics[width=\linewidth]{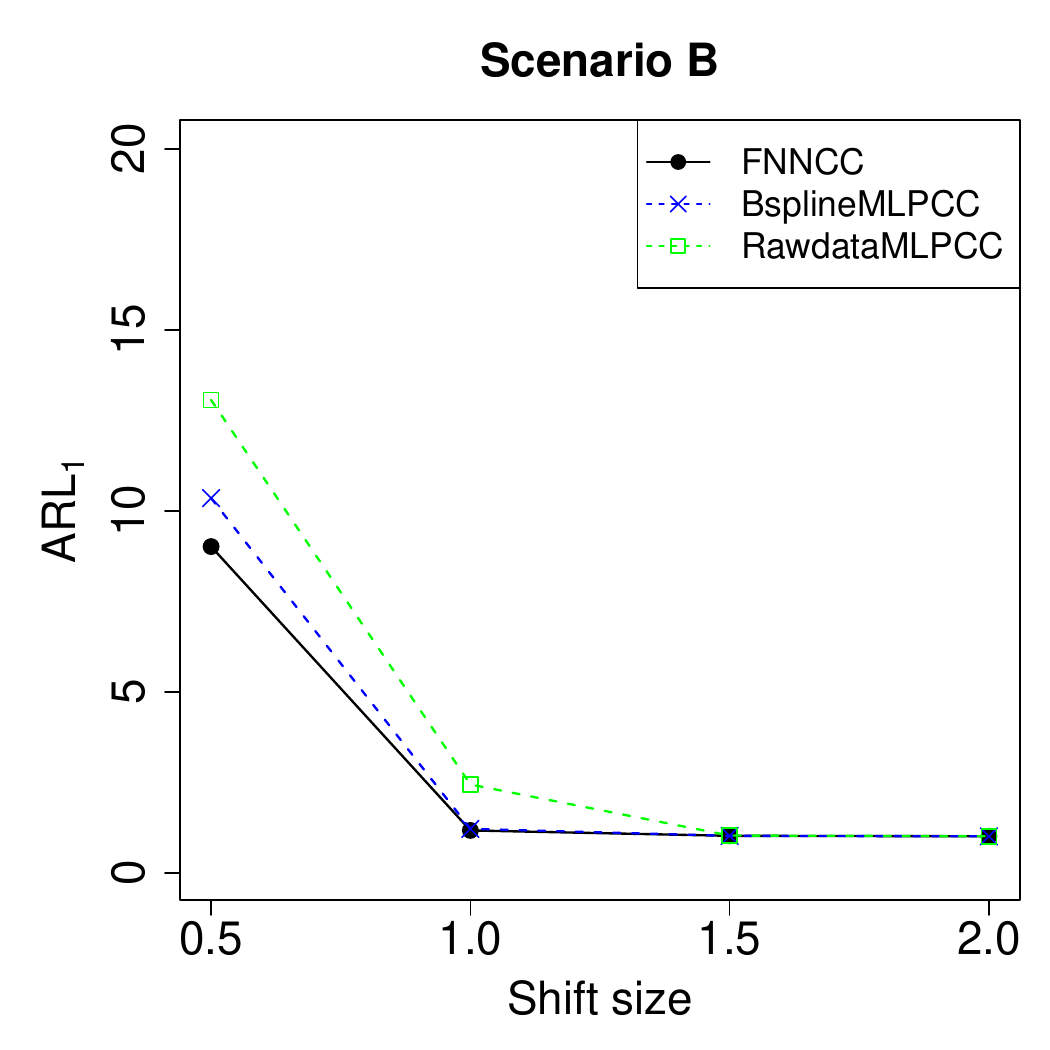}
\label{fig: FNNCC vs NNs exponential no cov shift}
\end{minipage}
\hspace{3mm} 
\begin{minipage}{0.41\textwidth}
\includegraphics[width=\linewidth]{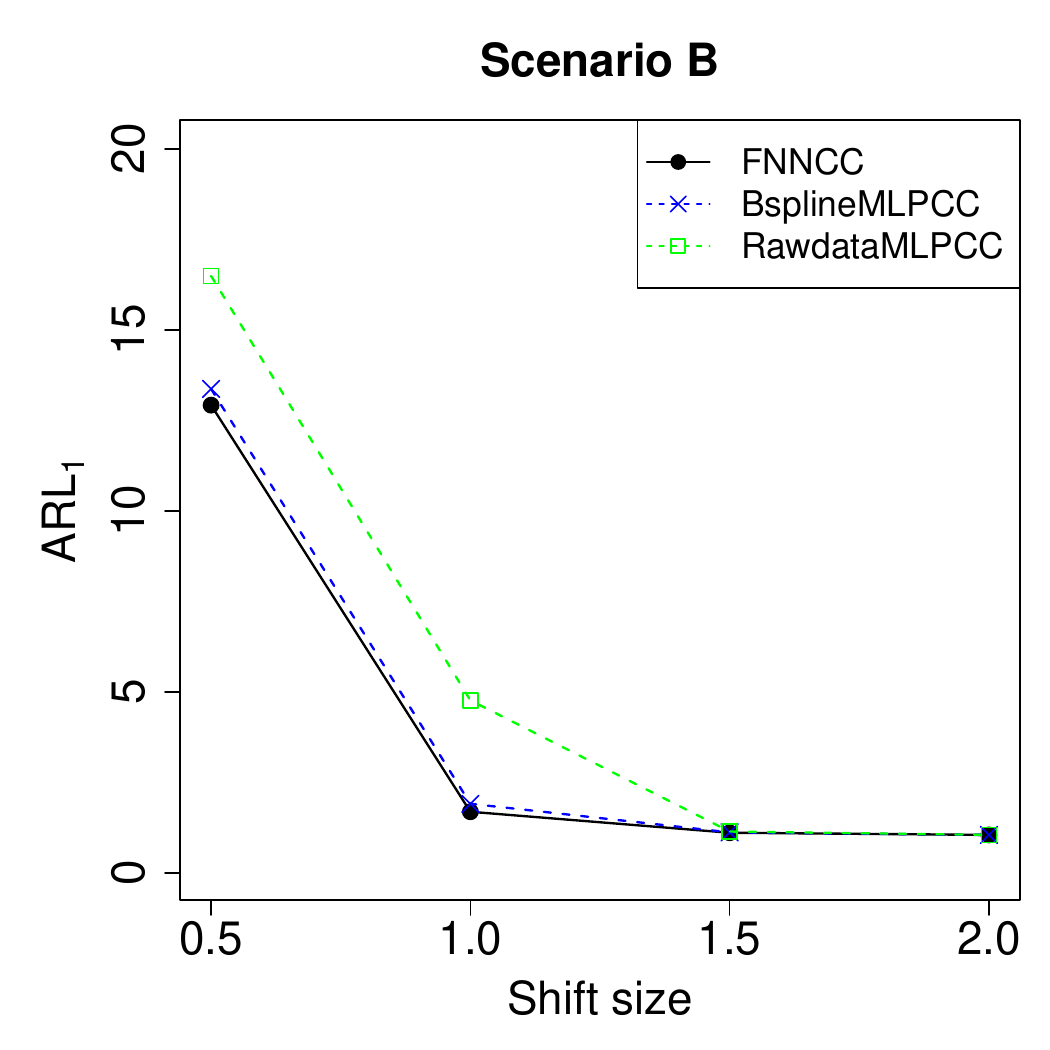}
\label{fig: FNNCC vs NNs exponential cov shift}
\end{minipage}
\caption{Estimated $ARL_1$ achieved by FNNCC, RawdataMLPCC, and BsplineMLPCC in Scenario B both in the presence (a) and absence (b) of a mean covariate shift, as a function of the mean shift size of the scalar response $\Delta \mu_y =  \{0.5,1,1.5,2 \} s^{y^*}$.}
\label{fig: FNNCC vs NNs exponential}
\end{figure}
%
\begin{figure}[hbt!]
\centering
\begin{minipage}{0.41\textwidth}
\includegraphics[width=\linewidth]{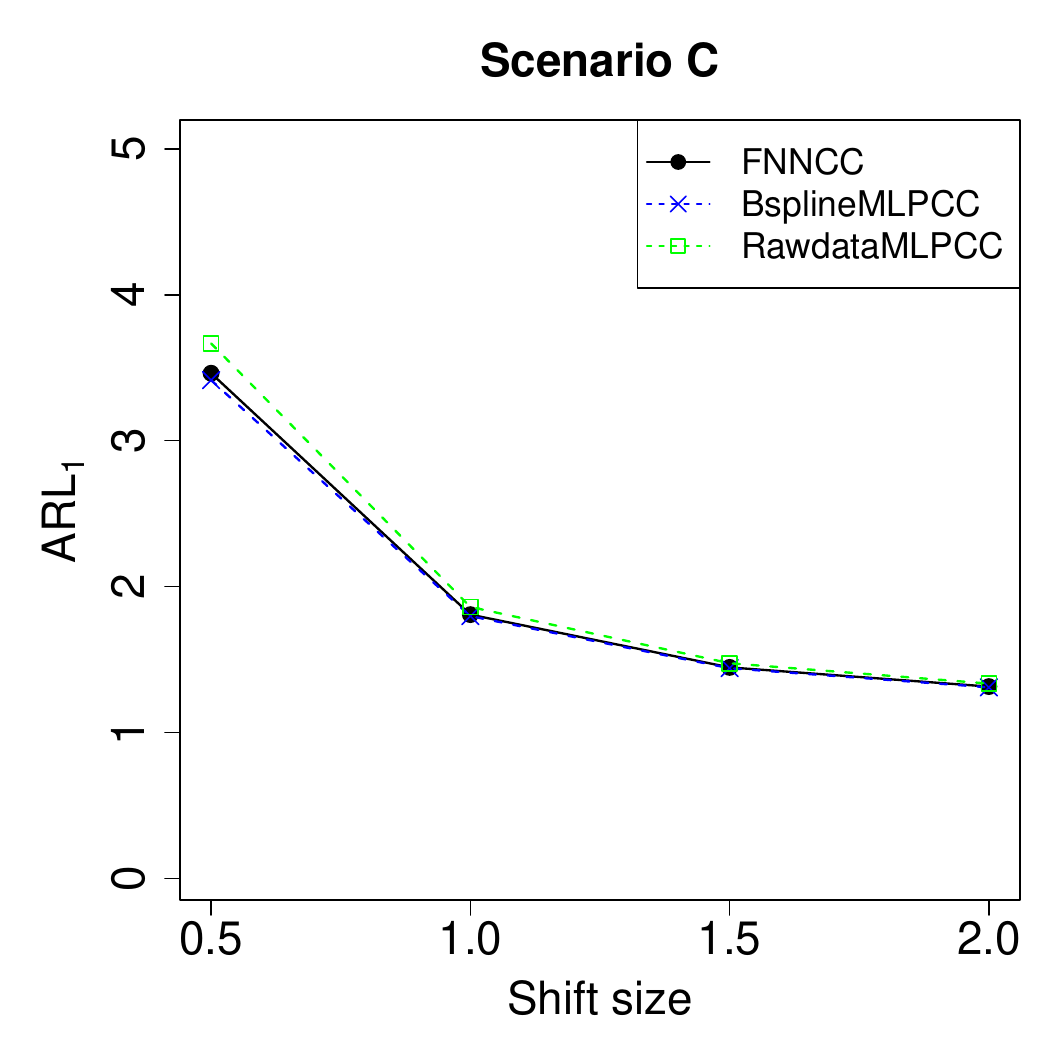}
\label{fig: FNNCC vs NNs abs no cov shift}
\end{minipage}
\hspace{3mm} 
\begin{minipage}{0.41\textwidth}
\includegraphics[width=\linewidth]{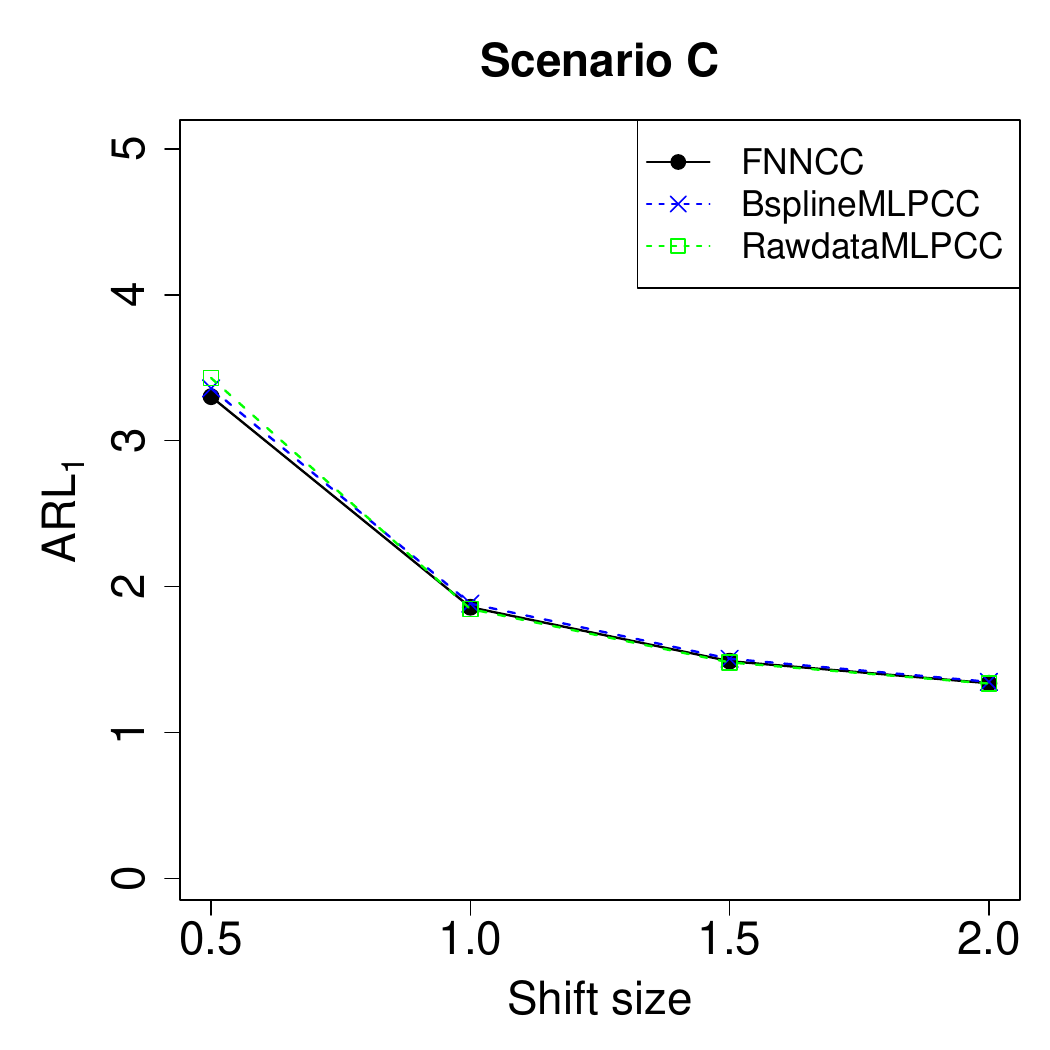}
\label{fig: FNNCC vs NNs abs cov shift}
\end{minipage}
\caption{Estimated $ARL_1$ achieved by FNNCC, RawdataMLPCC, and BsplineMLPCC in Scenario C both in the presence (a) and absence (b) of a mean covariate shift, as a function of the mean shift size of the scalar response $\Delta \mu_y = \{0.5,1,1.5,2 \} s^{y^*}$.}
\label{fig: FNNCC vs NNs abs}
\end{figure}
%
\begin{figure}[hbt!]
\centering
\begin{minipage}{0.41\textwidth}
\includegraphics[width=\linewidth]{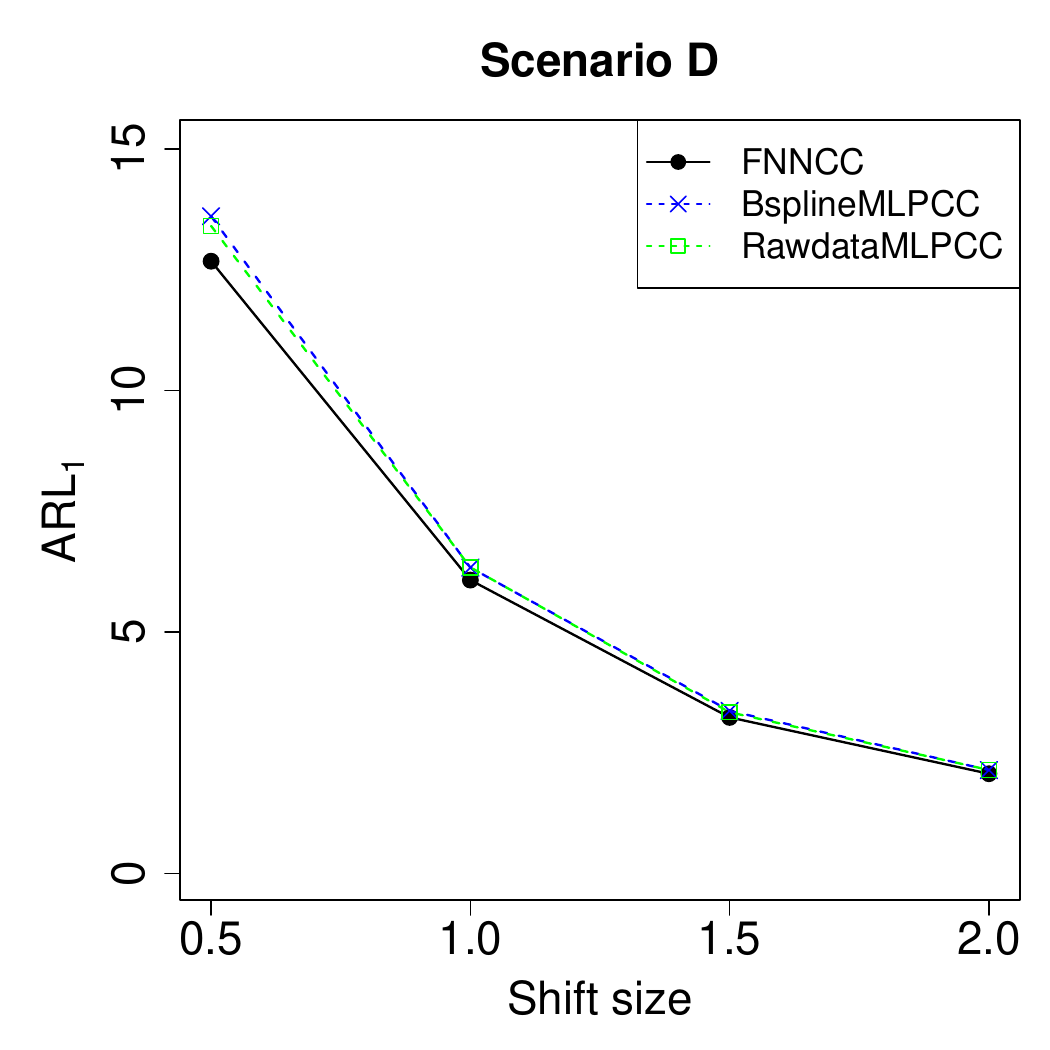}
\label{fig: FNNCC vs NNs log no cov shift}
\end{minipage}
\hspace{3mm} 
\begin{minipage}{0.41\textwidth}
\includegraphics[width=\linewidth]{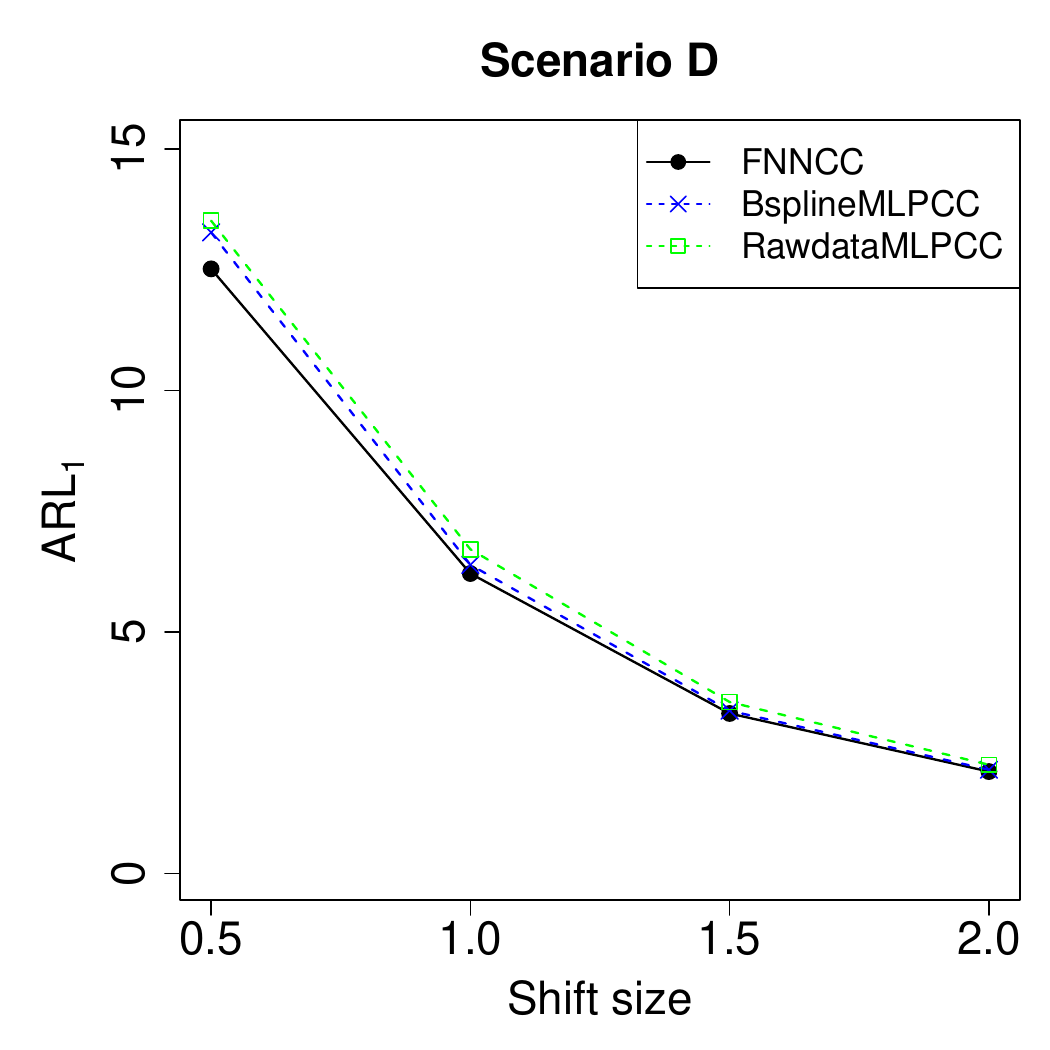}
\label{fig: FNNCC vs NNs log cov shift}
\end{minipage}
\caption{Estimated $ARL_1$ achieved by FNNCC, RawdataMLPCC, and BsplineMLPCC in Scenario D both in the presence (a) and absence (b) of a mean covariate shift, as a function of the mean shift size of the scalar response $\Delta \mu_y =  \{0.5,1,1.5,2 \} s^{y^*}$.}
\label{fig: FNNCC vs NNs log}
\end{figure}
%
\begin{figure}[hbt!]
\centering
\begin{minipage}{0.41\textwidth}
\includegraphics[width=\linewidth]{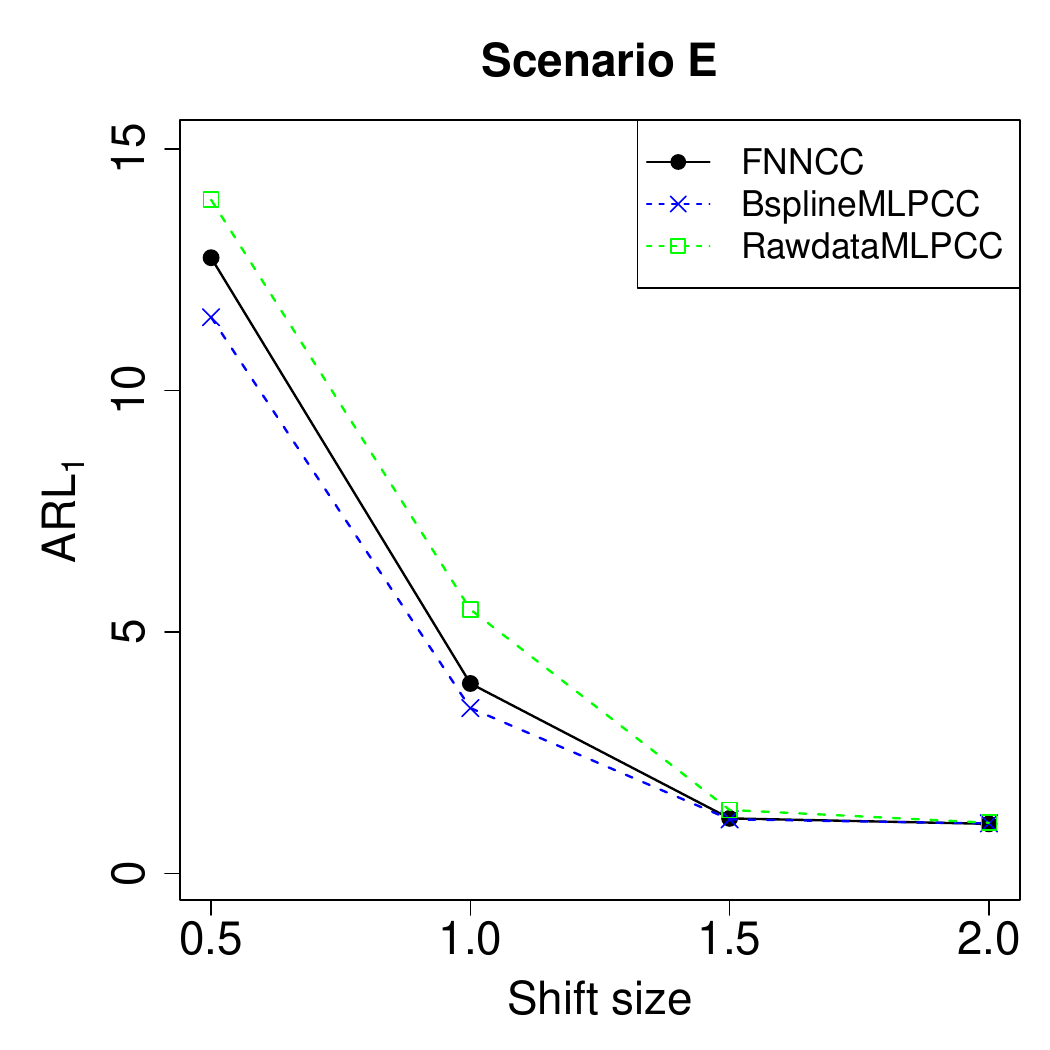}
\label{ffig: FNNCC vs NNs square no cov shift}
\end{minipage}
\hspace{3mm} 
\begin{minipage}{0.41\textwidth}
\includegraphics[width=\linewidth]{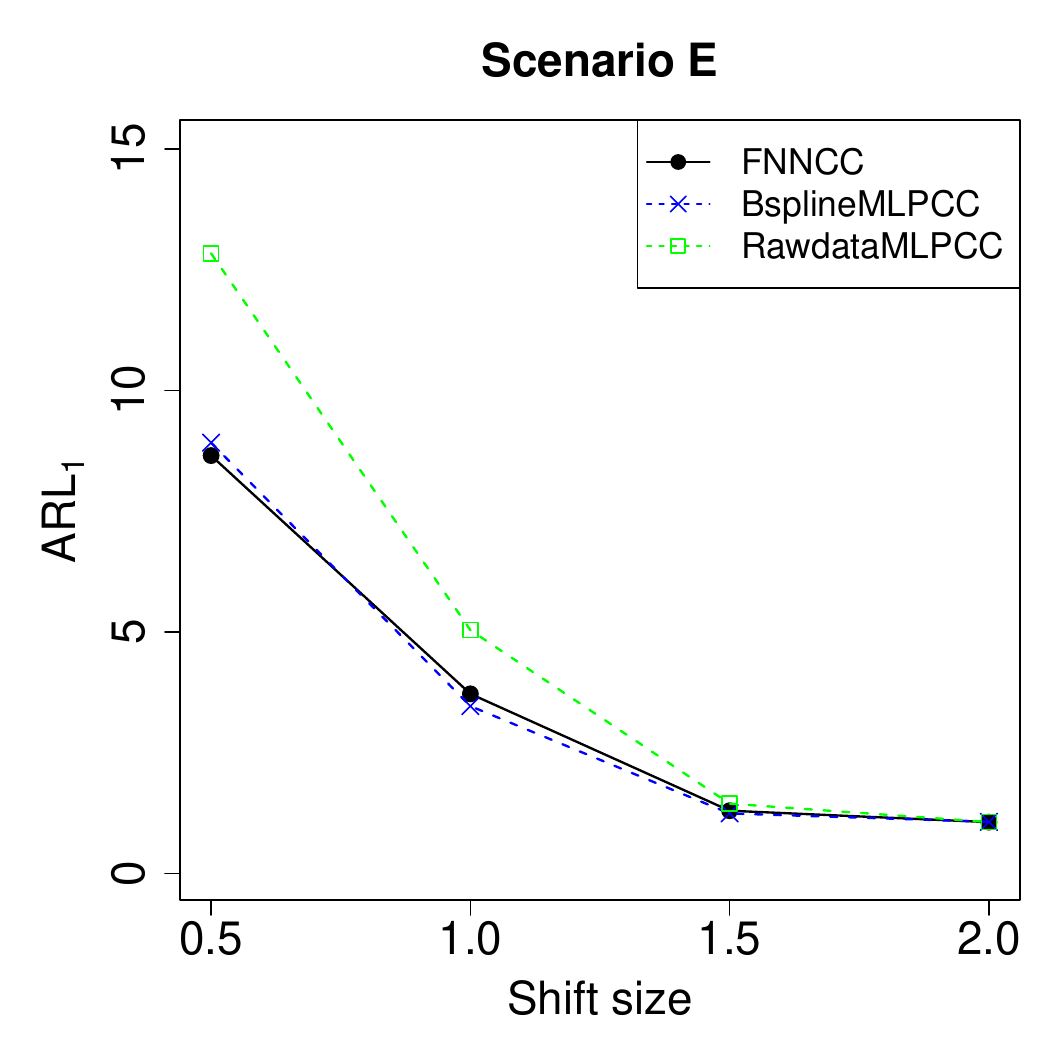}
\label{fig: FNNCC vs NNs square cov shift}
\end{minipage}
\caption{Estimated $ARL_1$ achieved by FNNCC, RawdataMLPCC, and BsplineMLPCC in Scenario E both in the presence (a) and absence (b) of a mean covariate shift, as a function of the mean shift size of the scalar response $\Delta \mu_y =  \{0.5,1,1.5,2 \} s^{y^*}$.}
\label{fig: FNNCC vs NNs square}
\end{figure}

\printcredits

\clearpage

\bibliographystyle{apalike}

\bibliography{cas-refs}





\end{document}